\documentclass{emulateapj}

\usepackage{graphicx}
\usepackage{natbib}
\usepackage[usenames]{color}
\usepackage{subfigure}
\usepackage{multirow}
\usepackage{amsmath}
\usepackage{xspace}
\usepackage{flafter}

\newcommand{\cl}{\ensuremath{C_{\ell}}\xspace}

\shorttitle{First Season QUIET Observations}
\shortauthors{The QUIET Collaboration}

\begin{document}

\setlength{\pdfpageheight}{\paperheight}
\setlength{\pdfpagewidth}{\paperwidth}

\title{First season QUIET observations: Measurements of CMB polarization power spectra at 43 GHz in the multipole range $25 \le \ell \le 475$}

\author{
QUIET Collaboration---C.~Bischoff\altaffilmark{1,22},
A.~Brizius\altaffilmark{1,2},
I.~Buder\altaffilmark{1},
Y.~Chinone\altaffilmark{3,4},
K.~Cleary\altaffilmark{5},
R.~N.~Dumoulin\altaffilmark{6},
A.~Kusaka\altaffilmark{1},
R.~Monsalve\altaffilmark{7},
S.~K.~N\ae ss\altaffilmark{8},
L.~B.~Newburgh\altaffilmark{6,23},
R.~Reeves\altaffilmark{5},
K.~M.~Smith\altaffilmark{1,23},
I.~K.~Wehus\altaffilmark{9},
J.~A.~Zuntz\altaffilmark{10,11,12},
J.~T.~L.~Zwart\altaffilmark{6},
L.~Bronfman\altaffilmark{13},
R.~Bustos\altaffilmark{7,13,14},
S.~E.~Church\altaffilmark{15},
C.~Dickinson\altaffilmark{16},
H.~K.~Eriksen\altaffilmark{8,17},
P.~G.~Ferreira\altaffilmark{10},
T.~Gaier\altaffilmark{18},
J.~O.~Gundersen\altaffilmark{7},
M.~Hasegawa\altaffilmark{3},
M.~Hazumi\altaffilmark{3},
K.~M.~Huffenberger\altaffilmark{7},
M.~E.~Jones\altaffilmark{10},
P.~Kangaslahti\altaffilmark{18},
D.~J.~Kapner\altaffilmark{1,24},
C.~R.~Lawrence\altaffilmark{18},
M.~Limon\altaffilmark{6},
J.~May\altaffilmark{13},
J.~J.~McMahon\altaffilmark{19},
A.~D.~Miller\altaffilmark{6},
H.~Nguyen\altaffilmark{20},
G.~W.~Nixon\altaffilmark{21},
T.~J.~Pearson\altaffilmark{5},
L.~Piccirillo\altaffilmark{16},
S.~J.~E.~Radford\altaffilmark{5},
A.~C.~S.~Readhead\altaffilmark{5},
J.~L.~Richards\altaffilmark{5},
D.~Samtleben\altaffilmark{2,25},
M.~Seiffert\altaffilmark{18},
M.~C.~Shepherd\altaffilmark{5},
S.~T.~Staggs\altaffilmark{21},
O.~Tajima\altaffilmark{1,3},
K.~L.~Thompson\altaffilmark{15},
K.~Vanderlinde\altaffilmark{1,26},
R.~Williamson\altaffilmark{6,27},
B.~Winstein\altaffilmark{1}
}

 \vspace{+0.2in}

\altaffiltext{1}{Kavli Institute for Cosmological Physics, Department of Physics, Enrico Fermi Institute, The University of Chicago, Chicago, IL 60637, USA; send correspondence to A.~Kusaka, \textbf{akito@kicp.uchicago.edu}}
\altaffiltext{2}{Max-Planck-Institut f\"ur Radioastronomie, Auf dem H\"ugel 69, 53121 Bonn, Germany}
\altaffiltext{3}{High Energy Accelerator Research Organization (KEK), 1-1 Oho, Tsukuba, Ibaraki 305-0801, Japan}
\altaffiltext{4}{Astronomical Institute, Graduate School of Science, Tohoku University, Aramaki, Aoba, Sendai 980-8578, Japan}
\altaffiltext{5}{Cahill Center for Astronomy and Astrophysics, California Institute of Technology, 1200 E. California Blvd M/C 249-17, Pasadena, CA 91125, USA}
\altaffiltext{6}{Department of Physics and Columbia Astrophysics Laboratory, Columbia University, New York, NY 10027, USA}
\altaffiltext{7}{Department of Physics, University of Miami, 1320 Campo Sano Drive, Coral Gables, FL 33146, USA}
\altaffiltext{8}{Institute of Theoretical Astrophysics, University of Oslo, P.O. Box 1029 Blindern, N-0315 Oslo, Norway}
\altaffiltext{9}{Department of Physics, University of Oslo, P.O. Box 1048 Blindern, N-0316 Oslo, Norway}
\altaffiltext{10}{Department of Astrophysics, University of Oxford, Keble Road, Oxford OX1 3RH, UK}
\altaffiltext{11}{Oxford Martin School, 34 Broad Street, Oxford OX1 3BD, UK}
\altaffiltext{12}{Department of Physics and Astronomy, University College London, Gower Street, London WC1E, UK}
\altaffiltext{13}{Departamento de Astronom\'ia, Universidad de Chile, Casilla 36-D, Santiago, Chile}
\altaffiltext{14}{Departamento de Astronom\'ia, Universidad de Concepci\'on, Casilla 160-C, Concepci\'on, Chile}
\altaffiltext{15}{Kavli Institute for Particle Astrophysics and Cosmology and Department of Physics, Stanford University, Varian Physics Building, 382 Via Pueblo Mall, Stanford, CA 94305, USA}
\altaffiltext{16}{Jodrell Bank Centre for Astrophysics, Alan Turing Building, School of Physics and Astronomy, The University of Manchester, Oxford Road, Manchester M13 9PL, UK}
\altaffiltext{17}{Centre of Mathematics for Applications, University of Oslo, P.O. Box 1053 Blindern, N-0316 Oslo, Norway}
\altaffiltext{18}{Jet Propulsion Laboratory, California Institute of Technology, 4800 Oak Grove Drive, Pasadena, CA, USA 91109}
\altaffiltext{19}{Department of Physics, University of Michigan, 450 Church Street, Ann Arbor, MI 48109, USA}
\altaffiltext{20}{Fermi National Accelerator Laboratory, Batavia, IL 60510, USA}
\altaffiltext{21}{Joseph Henry Laboratories of Physics, Jadwin Hall, Princeton University, Princeton, NJ 08544, USA}
\altaffiltext{22}{Current address: Harvard-Smithsonian Center for Astrophysics, 60 Garden Street MS 43, Cambridge, MA 02138, USA}
\altaffiltext{23}{Current address: Joseph Henry Laboratories of Physics, Jadwin Hall, Princeton University, Princeton, NJ 08544, USA}
\altaffiltext{24}{Current address: Micro Encoder Inc., Kirkland, WA 98034, USA}
\altaffiltext{25}{Current address: Nikhef, Science Park, Amsterdam, The Netherlands}
\altaffiltext{26}{Current address: Department of Physics, McGill University, 3600 Rue University, Montreal, Quebec H3A 2T8, Canada}
\altaffiltext{27}{Current address: Kavli Institute for Cosmological Physics, Enrico Fermi Institute, The University of Chicago, Chicago, IL 60637, USA}

\slugcomment{
This paper should be cited as ``QUIET Collaboration et al. (2011)''
}
\journalinfo{Published in ApJ 741 111 (2011)}

\begin{abstract}
The Q/U Imaging ExperimenT (QUIET) 
employs coherent receivers at 43\,GHz and 95\,GHz,  operating on the Chajnantor plateau in the Atacama Desert in Chile,  to measure the anisotropy in the polarization of the CMB. 
 QUIET primarily targets the B modes from primordial gravitational waves. The combination of these frequencies gives  sensitivity to foreground contributions from diffuse Galactic synchrotron radiation.   
Between 2008 October and 2010 December, over 10,000\,hours of data were collected, first with the 19-element 43-GHz array (3458\,hours) and then with the 90-element 95-GHz array.  Each array observes the same four fields, selected for low foregrounds, together covering $\approx1000$ square degrees.  This paper reports initial results from the 43-GHz receiver which has an array sensitivity to CMB fluctuations of   69\,$\mu$K$\sqrt{\text{s}}$. 
The data were extensively studied with a large suite of null tests before the power spectra, determined with two independent pipelines, were examined.  
Analysis choices, including data selection, were modified until the null tests passed.  Cross correlating maps with different telescope pointings is used to eliminate a bias.
This paper reports the EE,  BB, and EB power spectra in the multipole range $\ell=25$--475. 
With the exception of the lowest multipole bin for one of the fields, where a polarized foreground, consistent with Galactic synchrotron radiation,  is detected with $3$-$\sigma$ significance, the E-mode spectrum is consistent with the
$\Lambda$CDM model, confirming the only previous detection of the first acoustic peak.  The B-mode spectrum is consistent with zero, leading to a measurement of the tensor-to-scalar ratio of $r=0.35^{+1.06}_{-0.87}$.  The combination of a new time-stream ``double-demodulation" technique, side-fed Dragonian optics, natural sky rotation, and frequent boresight rotation leads to the lowest level of systematic contamination in the B-mode power so far reported, below the level of $r = 0.1$.  

 \vspace{+0.2in}

\end{abstract}


\maketitle

\nocite{Chiang:2010}

\section{Introduction}
\label{sec:intro}

The inflationary paradigm resolves several outstanding
issues in cosmology, including the flatness, horizon, and monopole
problems, and it provides a compelling explanation for the origin of
structure in the Universe \citep[e.g.][and references therein]{liddle:2000}. So far all cosmological data, including measurements of Cosmic Microwave Background (CMB) anisotropies,  support this paradigm; still the underlying fundamental physics responsible for inflation is unknown.  
Inflation produces a stochastic background of
gravity waves that induce odd-parity tensor ``B modes'' at large
angular scales in the CMB polarization. 
If these primordial B modes, parametrized by the tensor-to-scalar
ratio $r$,  are detected, one can learn about the
energy scale of inflation. In many attractive slow-roll models, this scale is given approximately 
by $  r^{1/4}\times 10^{16}$\,GeV.  For large-field models, the energy scale is near the Grand Unification Scale in particle physics, so that $r \gtrsim 0.01$.
A new generation of experiments aims for good sensitivity in this range of $r$.  Establishing the existence of primordial B modes would both verify an important prediction of inflation and provide access to physics at an incredibly high energy scale.

The most
stringent limit to date is $r < 0.20$ at the $95\%$ confidence
level \cite[]{Komatsu:2010fb} set by a combination of CMB--temperature-anisotropy measurements, baryon acoustic oscillations, and supernova
observations, but cosmic variance prohibits improvements using only these measurements.

E-mode polarization has now been detected by many
experiments \citep[e.g.,][]{kovac:2002, leitch:2005, montroy:2006,
sievers:2007, Wu2007, bischoff:2008, larson:2010}.  These measurements are consistent with predictions 
from CMB--temperature-anisotropy measurements, and they provide new information on the epoch of reionization. Only BICEP has accurately measured 
 E-mode polarization in the region of the first acoustic peak
\cite[]{Chiang:2010}; that paper also reports the best limit on $r$ coming from cosmological B modes: $r < 0.72$ at the $95\%$ confidence level.
 
Experiments measuring B-mode polarization in the CMB
should yield the best information on  $r$, but this
technique is still in its infancy. 
B modes are expected to be at least an order of
magnitude smaller than the E modes so  control
of systematic errors and foregrounds  will be particularly critical. Below $\approx 90$\,GHz, the dominant foreground 
comes from Galactic synchrotron emission, while at higher frequencies, emission from thermal dust dominates. Most planned or operating CMB polarization
experiments employ bolometric detectors observing most comfortably
 at frequencies $\gtrsim 90$\,GHz, so they cannot estimate synchrotron contamination from their own data. 
 
 The Q/U Imaging ExperimenT (QUIET) is one of two CMB
polarization experiments to observe at frequencies suitable for
addressing synchrotron contamination, making observations at 43\,GHz (Q band) and 95\,GHz
(W band) and with sufficient
sensitivity to begin to probe primordial B modes.  The other is \textit{Planck} \citep{Tauber2010}.

QUIET uses
compact polarization-sensitive  modules based upon High--Electron-Mobility Transistor (HEMT) amplifiers, combined with a new time-stream ``double-demodulation'' technique,
side-fed Dragonian optics (for the first time in a CMB polarization
experiment), natural sky rotation, and frequent rotation about the optical axis to
achieve a very low level of  contamination in the multipole range where a primordial--B-mode signal is expected.

Between 2008 October and 2010 December, QUIET collected over 10,000\,hours of data, split between the Q-band and W-band receivers.  Here we report first results from the first season of 3458\,hours of Q-band observation. 
The principal investigator for QUIET was our recently deceased
colleague, Bruce Winstein, whose intellectual and scientific guidance
were crucial to QUIET in all its stages, from design through analysis,
through writing this paper.

After describing the instrument, observations, and detector calibrations (Sections 2, 3, and 4), we discuss our analysis techniques and consistency checks (5 and 6).  CMB power spectra are then presented together with a foreground detection (7).  We evaluate our systematic errors (8) and then conclude (9).

\section{The Instrument}
\label{sec:instrument}

\begin{figure}[htbp]
\centering
\includegraphics[width=3.5in]{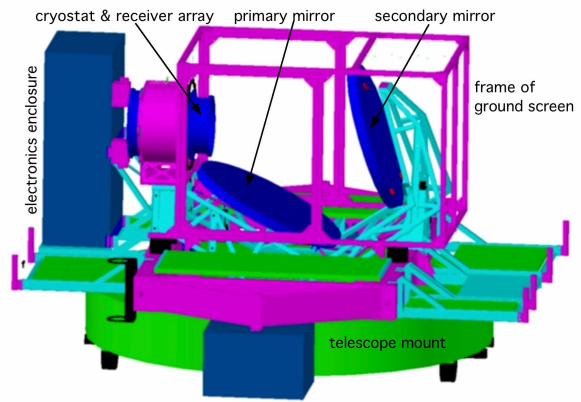}
\caption{Overview of the QUIET instrument.  The cryostat and 1.4-m telescope mirrors are enclosed in a rectangular comoving absorbing ground screen;  in this figure its walls are transparent.  The telescope, cryostat and electronics are mounted on a single platform attached to the deck bearing, which allows rotations around the instrument's optical axis.
\label{fig:instrument_overview}}
\end{figure}

The QUIET instrument comprises an array of correlation polarimeters cooled to 20\,K and coupled to a dual-reflector telescope, installed on a three-axis mount inside a comoving ground screen.  The instrument is illustrated in Figure~\ref{fig:instrument_overview}.  Further details are given below and in \cite{Lauraproc}, \cite{Akitoproc}, and \cite{buder:77411D}.

The Q-band QUIET receiver
is a 19-element array containing 17  low-noise correlation polarimeters, each   
 simultaneously measuring the Stokes Q, U, and I parameters, and two 
CMB differential-temperature monitors. 

QUIET uses a 1.4-m
classical side-fed Dragonian antenna~\citep{Dragone}. This
consists of a parabolic primary, a concave hyperbolic secondary along with a
platelet array of corrugated feed horns~\citep{Gunder}. These elements
are oriented in a way
to satisfy the Mizuguchi condition~\citep{Mizuguchi} in order to minimize
cross polar response, and unlike dual offset classical Gregorian or Cassegrain
antennas, the elements combine to generate high gain with low sidelobe
response over a wide field of view~\citep{2004ITAP...52...12C}.
The telescope is described in detail in \cite{Imbriale}.
It yields a
full-width half-maximum (FWHM) beam size of $27\farcm3$ and a roughly circular field of view of 7$^{\circ}$ diameter.  Radiation from each feed horn enters a septum polarizer \citep{Bornemann} which separates left and right circularly-polarized components ($L$ and $R$) into two waveguide ports which mate to a QUIET correlation module, detailed below.  

The module array and feed horns are cooled to 20\,K in a cryostat to reduce instrumental noise.  An electronics enclosure mounted next to the cryostat houses the electronics necessary for biasing the modules and recording their data.
The cryostat, electronics, and telescope are  installed on the former CBI mount  
\citep{Padin:2001df}. This mount provides three-axis motion:  azimuth, elevation, and rotation about the optical axis.  This last is called ``deck'' rotation.  

The cryostat and telescope are enclosed by an absorbing
comoving ground screen.
The ground screen was designed to have two parts, but the upper section (not shown in Fig.~\ref{fig:instrument_overview})  was not installed until after the Q-band instrument was removed.  Its absence was correctly anticipated to result in two far sidelobes, which were mapped with a high-power source by  the QUIET W-band instrument in the field and measured to be $\lesssim -60$\,dB with the QUIET Q-band instrument when the Sun passed through them.  
The effects of these sidelobes are mitigated through filtering and data selection (Sections~~\ref{sec:data:tod:filter} and \ref{subsec:cuts}). 
 Section~\ref{sec:systematics_sun} shows that any residual contamination is small.

Each QUIET Q-band correlation module, in a footprint of only $5.1\times 5.1\,\mbox{cm}^2$, receives  the circular polarization modes of the celestial radiation and outputs Stokes $Q$, $U$ and $I$ as follows.  Each input is independently amplified and passed through a phase switch.  One phase switch alternates  the sign of the signal voltage at 4\,kHz, while the other switches at 50\,Hz.  The two signals are combined in a  $180^\circ$ hybrid coupler, with outputs proportional to the sum and difference of the inputs.  Since the module inputs are proportional to $(L, R)=(E_{x}\pm iE_{y})/ \sqrt{2}$, where $E_x$ and $E_y$ are orthogonal components of the incident electric  field,  the coupler outputs are amplified versions of $E_x$ and $iE_y$, with the phase switch reversing their roles.  Half of each output is bandpass filtered and rectified by  a pair of detector diodes, while the other half passes into  a $90^\circ$ hybrid coupler. A second pair of bandpass filters and detector diodes measures the power from this coupler's outputs \citep{kangaslahti2006}.  

Synchronous demodulation of the 4-kHz phase switching yields measurements 
of Stokes $+Q$ and $-Q$ on the first two diodes and Stokes $+U$ and $-U$  on the remaining two. This high-frequency differencing suppresses low-frequency atmospheric fluctuations as well as  $ 1/f$ noise from the amplifiers, detector diodes, bias electronics, and data-acquisition electronics. 
Subsequent demodulation of the 50-Hz phase switching removes spurious instrumental polarization generated by unequal transmission coefficients in the phase-switch circuits.  The resulting   four ``double-demodulated'' time streams are the  polarization channels.

Averaging the output of each diode rather than demodulating it results  
in a measurement of Stokes $I$, hereafter called total power, denoted ``TP.''  
The TP time streams are useful for monitoring the weather and the stability of the detector responsivities, but suffer too much contamination from $1/f$ noise to 
constrain the CMB temperature anisotropy. Therefore, the Q-band instrument 
includes two  correlation modules that are coupled to a pair of neighboring feed horns to measure the temperature difference between them, in a scheme similar to the \textit{WMAP} differencing assemblies \citep{Jarosik2003}. These differential-temperature modules provide calibration data for the telescope pointing, beams, and sidelobes, as well as CMB data.
Their feed horns are in the outer ring of the close-packed hexagonal array, $\approx3^\circ$ from the center.

Here we summarize several array-wide characteristics of the polarimeters. 
Bandpass measurements in the lab and at the start of the observing season find that the average center frequency is $43.1\pm 0.4$\,GHz,  and the average bandwidth is $7.6\pm 0.5$\,GHz. We calculate the noise power spectra of the double-demodulated polarimeter time streams from each 40--90-minute observation to assess their $1/f$ knee frequencies and white-noise levels (see Section~\ref{subsec:TOD}).  
The median $1/f$ knee frequency is 5.5\,mHz, well below the telescope scan frequencies of 45--100\,mHz.

From the white-noise levels and responsivities 
(Section~\ref{sec:calibration_responsivity})
 we find an array sensitivity\footnote{This is the sensitivity for 62 polarization channels.  Six of 68 polarization channels are non-functional---an array yield of 92\%. }
to CMB fluctuations of 69\,$\mu$K$\sqrt{\text{s}}$,
 such that the mean polarized sensitivity per module is   
 280\,$\mu\mbox{K}\sqrt{\text{s}}$.

\begin{figure}[phtb]
 \begin{center}
    \includegraphics[width=1.0\linewidth]{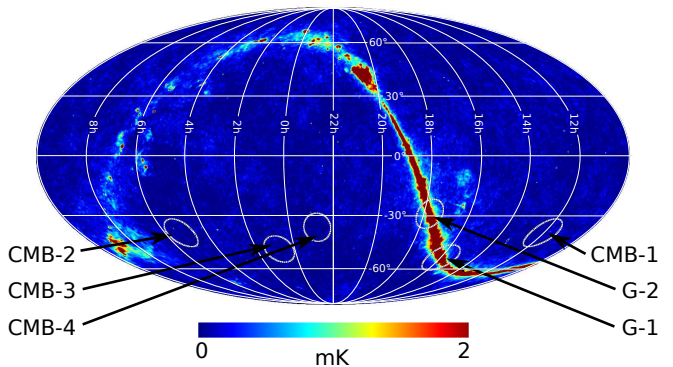}
  \end{center}

 \vspace{-0.15in}

  \caption{\label{fig:quiet_patches} The CMB and Galactic patches, in
    equatorial coordinates, superimposed on a Q-band all-sky
    \textit{WMAP} 7-year temperature map \citep{Jarosik:2010iu}. Note
    that the Galactic-plane temperature signal saturates the color scale. Patch G-2 is the Galactic
    center. } 
\vspace{1mm}
\end{figure}

\section{Observations}
\label{sec:observations}

QUIET is located on the Chajnantor plateau in the Atacama Desert of
northern Chile ($67\degr 45\arcmin 42\arcsec$\,W, $23\degr 01\arcmin
42\arcsec$\,S). A combination of high altitude (5080\,m) and extreme
dryness results in excellent observing conditions for most of the year.
During the eight months of QUIET Q-band observations, the median precipitable water vapor (PWV) measured at the nearby APEX site \citep{gusten:2006} was 1.2\,mm.

We began observations with the Q-band receiver on 
2008~October~24, and took 3458\,hours of data until 2009~June~13 (when the
receiver was replaced on the telescope by the 90-element W-band
receiver).  Of these data,  77\% are for CMB, 
with 12\% of the observing time used for Galactic fields,  7\% for calibration
sources, and 4\% cut due to obvious instrumental problems such as lack of telescope motion. We observe 24\,hours a day, except when interrupted.  Our full-season operating efficiency is  63\%;  causes of downtime include occasional snow, power outages, and mechanical failures.

\begin{table}[htdp]
\begin{center}
\caption{Patch Locations and Integration Times}

 \vspace{-0.05in}

\begin{tabular}{cccc}
\hline
\hline
Patch & RA & Dec. & Integration \\
&   \multicolumn{2}{c}{(J2000)}  & Hours \\ 
\hline
CMB-1 & $12^h 04^m $ & $-39\degr 00\arcmin$ & $905$ \\ 
CMB-2 & $05^h 12^m $ & $-39\degr 00\arcmin$ & $703$ \\ 
CMB-3 & $00^h 48^m $ & $-48\degr 00\arcmin$ & $837$ \\ 
CMB-4 & $22^h 44^m $ & $-36\degr 00\arcmin$ & $223$ \\ 
\hline 
G-1 & $16^h 00^m $ & $-53\degr 00\arcmin$ & $311$ \\
G-2 & $17^h 46^m $ & $-28\degr 56\arcmin$ & $92$ \\
\hline
\end{tabular}
\label{tbl:quiet_patches}
\\
\end{center}

 \vspace{-0.1in}

\tablecomments{The central equatorial coordinates and 
integration times for each observing patch. 
G-1 and G-2 are Galactic patches.}
\end{table}

\subsection{Field Selection}
\label{obs:field-selection}

We observe four CMB fields, referred to henceforth as ``patches.'' Table \ref{tbl:quiet_patches} lists their
center positions and total integration times, while
Figure \ref{fig:quiet_patches} indicates their positions on the
sky\footnote{Patch CMB-3 partially overlaps with the field the
BICEP collaboration has observed for CMB analysis~\cite[]{Chiang:2010}. The data may
be used for future analysis cross-correlating maps from the two experiments.}.
The number of patches is determined by the requirement 
to always have 
one patch above the lower elevation limit of the mount ($43^\circ$). 
The specific
positions of each patch were chosen to minimize foreground emission using \textit{WMAP} 3-year data. The area of each patch is $\approx
250\,\textrm{deg}^2$. In addition to the four CMB patches, we observe
two Galactic patches. These allow us to constrain the spectral
properties of the polarized low-frequency foregrounds with a high
signal-to-noise ratio. The results from the Galactic observations will be
presented in a future publication.

\subsection{Observing Strategy}
\label{obs:strategy}

Scanning  the telescope modulates the signal from the sky,
converting CMB angular scales into frequencies in the polarimeter time streams.
Since QUIET targets large angular scales, fast scanning ($\approx5\degr$\,s$^{-1}$ in azimuth) 
is critical to ensuring that the polarization modes of interest
appear at higher frequencies than the atmospheric and instrumental $1/f$ knee frequencies.

So that each module sees a roughly-constant atmospheric signal,
QUIET scans are
periodic motions solely in azimuth with both the elevation and 
deck-rotation axes fixed. 
Each scan has an amplitude of $7.5\degr$ on the sky, 
with period  10--22$\,\textrm{s}$.  These azimuth scans are repeated for 40--90\,minutes;
each set of scans at fixed elevation is denoted a ``constant-elevation scan'' (CES).
We repoint the telescope and begin a new CES when the patch center has moved by $15\degr$
in order to build up data over an area of $\approx15\degr\times15\degr$ for each patch.
Note that a central region $\simeq 8\degr$ across is observed by all polarimeters since the instrument's field of view has a diameter of $\simeq\, 7\degr$.
Diurnal sky rotation and weekly deck rotations provide uniform parallactic-angle coverage of the patch, and ensure that its  peripheral regions are also observed by multiple polarimeters.

\begin{table}[htdp]
        \begin{center}
         \caption{Regular Calibration Observations}
	\label{tab:calibration_observations}

 \vspace{-0.05in}

        \begin{tabular}{l c c } 
        \hline \hline
        Source   & Schedule & Duration (min.) \\
        \hline
        sky dips & every 1.5\,hours & ~3\\
        Tau~A    & every 1--2\,days  & 20 \\
        Moon     & weekly          & 60  \\
        Jupiter  & weekly          & 20  \\
        Venus    & weekly          & 20  \\
        RCW38    & weekly          & 20  \\
        \hline
        \end{tabular}
      
	\end{center}
\end{table}

\section{Calibration}
\label{sec:calibration}

Four quantities are required to convert polarimeter time streams into polarization power spectra: 
detector responsivities, a pointing model, detector polarization angles, and beam profiles.  To this end, a suite of calibration observations is performed throughout the season using astronomical sources (Taurus A--hereafter Tau~A, Jupiter, Venus, RCW38, and the Moon);   atmospheric measurements (``sky dips,'' which typically consist of three  elevation nods of $\pm 3^{\circ}$);  and instrumental sources (a rotating sparse wire grid and a polarized broadband noise source).  
From these we also measure instrumental polarization, as described below.
QUIET's regular calibration observations are summarized in Table~\ref{tab:calibration_observations}.

We typically use two or more methods to determine a calibration constant, taking the spread among the methods as an indication of the uncertainty. We show in Section~\ref{sec:systematics} that aside from the case of absolute responsivity, all calibration uncertainties lead to estimates of systematic effects on the power spectra well below statistical errors.  This  immunity comes from having a large number of detectors and highly-crosslinked polarization maps.

\begin{figure}[htbp]
\centering
\includegraphics[width=2.6in]{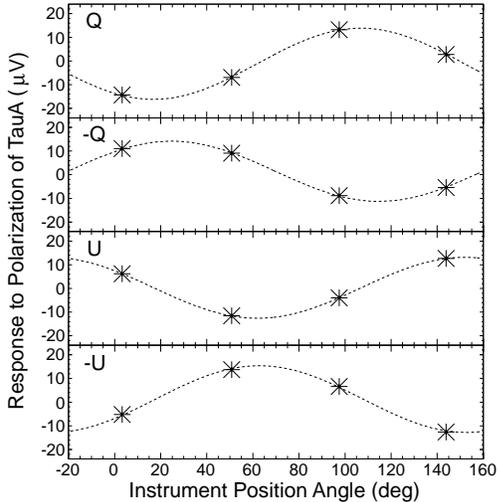}

 \vspace{-0.05in}

\caption{Polarimeter responses from the central feed horn to the polarization of Tau~A at four deck angles.  The horizontal axis corresponds to the position angle of the receiver focal plane in equatorial coordinates.  These data were collected with one correlation module in about 20\,minutes.  The errors are smaller than the points. From top to bottom, responses are shown for the detector diodes sensitive to the Stokes parameters $+$Q, $-$Q, $+$U, and $-$U, respectively.  For each, the fitted model is plotted as a dashed line.
    }
\label{fig:calibration_taua}
\end{figure}

\subsection{Responsivity}
\label{sec:calibration_responsivity}

The polarized flux from Tau~A provides a 5\,mK signal which we observe at four parallactic angles. The sinusoidal modulation of the signal induced by the changing parallactic angles is fitted to yield responsivity coefficients for each detector.  
Figure~\ref{fig:calibration_taua} 
shows the response of the four polarization channels from the central feed horn to Tau~A.  A typical responsivity is  2.3\,mV~K$^{-1}$,  with a precision from a single set of observations of 6\%.   
The absolute responsivity from Tau~A was measured most frequently
for the central feed horn. We choose its +Q diode detector to provide the fiducial absolute responsivity.

The  responsivities of other detectors relative to the fiducial detector are determined with the sky  dips as described below.   We have three independent means of assessing the relative responsivities among polarimeters:  from nearly-simultaneous measurements of the Moon, from simultaneous measurements of responses to the rotating sparse wire grid in post-season tests, and from Tau~A measurements.   The errors from these methods are 4\%, 2\%, and 6\% respectively, while the error from the sky-dip method is 4\%.  All the methods agree within errors.

Sky dips generate temperature signals of several 100\,mK and thus permit measurement of the TP responsivities.  The signals vary slightly with PWV.  We estimate the slope from the data as 4\%~mm$^{-1}$ and correct for it.  This slope is consistent  with the atmospheric model of \cite{Pardo2001}.  Because  the ratios of  the responsivities for the TP and polarized signals from each detector diode are stable quantities within a few percent of  unity, we use sky dips performed immediately before each CES to correct  short-term variations in the polarimeter responsivities.  The responsivities vary by  $\lesssim 10$\% over the course of a day,  due to changing thermal conditions for the bias electronics.
Further post-season tests  provide a physical model:   the relevant temperatures are varied intentionally while the responsivities are measured with sky dips.  We 
confirm  the results with the polarized broadband source.

We bound the uncertainty in the absolute responsivity of the polarimeter array at 6\%.  The largest contributions to this estimate are uncertainties in (1) the beam solid angle (4\%, see below), (2) the response difference between polarized and TP signals for each diode detector (3\%),  and (3) the Tau~A flux \citep[3\%,][]{Weiland2010}.  The first enters in converting the flux of Tau~A into $\mu$K, while the second enters because although one fiducial diode detector is calibrated directly from Tau~A, for the rest we find relative responsivities from sky dips and normalize by the  fiducial diode's responsivity.

For the differential-temperature modules, all detectors observe the signal from Jupiter simultaneously, providing the absolute responsivity for all channels upon comparison with the Jupiter flux from \cite{Weiland2010}. Observations of Venus \citep{VSA} and RCW38 agree with the Jupiter measurements within errors, 
and sky dips track short-term variations. We calibrate the absolute responsivity with 5\% accuracy.

\begin{figure}[htbp]
\centering
\includegraphics[width=1.8in]{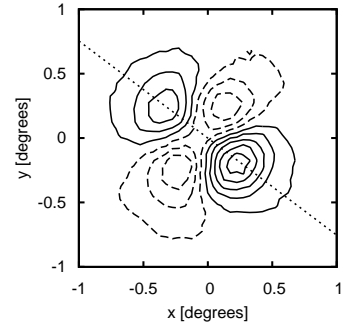}
\caption{Map of the polarization of the Moon from one detector diode. The amplitude of the quadrupole polarization visible here is $\approx 400$\,mK.   Similar maps are produced for all 17 polarization modules in the array with a single $\approx$~hour-long observation. The dotted line indicates the polarization orientation of the detector. Contours are spaced at intervals of 100\,mK, with negative contours indicated by dashed lines.
         }
\label{fig:calibration_moon_pol_map}
\end{figure}

 \vspace{0.5in}

\subsection{Pointing}
\label{sec:pointing}

The global pointing solution derives from a physical model of the
3-axis mount and telescope tied to observations of the Moon with the
central feed horn in the array, as well as Jupiter and Venus with the
differential-temperature feed horns.  Optical observations are taken
regularly with a co-aligned star camera and used to monitor the time
evolution of the pointing model. 

During the first two months in the season, a mechanical problem with
the deck-angle encoder results in pointing shifts.  The problem was
subsequently repaired. Based on pointing observations of the Moon and
other astronomical sources, we verify that these encoder shifts are less than $2^\circ$.  Systematic uncertainties induced by this problem are discussed in Section~\ref{sec:beam_sys}.

After the deck-angle problem is fixed, no significant evolution of the pointing model is found.  The difference in the mean pointing solution between the start and the end of the season is smaller than $1^\prime.$  Observations of the Moon and
Jupiter also provide the relative pointing among the feed horns.  The
root mean square (RMS) pointing error in the maps is $ 3\farcm5$.

\subsection{Detector Polarization Angles}
\label{sec:cal_pol_angle}

Our primary measurement of the polarization angle for each detector comes
from observing the radial polarization of the Moon,
as illustrated in Figure~\ref{fig:calibration_moon_pol_map}.
The measurement has a high signal-to-noise ratio and its inaccuracy is
dominated by systematic error due to the temperature gradient of the
Moon surface.  One can see the effect in the different amplitudes of the
two positive envelopes in Figure~\ref{fig:calibration_moon_pol_map}.
The fluctuations of the detector polarization-angle measurements over
many observations with different phases of the Moon and telescope
orientations are typically $1\degr$ in RMS.
Although simulations
suggest these fluctuations can be due to the failure to account in
analysis for the temperature gradient, we conservatively assign them as
upper limits on the fluctuations of the polarization angles during the
season.
Based
on this conservative limit, we estimate the systematic error in the
CMB--power-spectra measurement in Section~\ref{sec:gain_sys}, resulting
in a negligible contribution.

Two other less precise methods also give estimates of the detector angles:
fits to the Tau~A data, and determination of the phases of the
sinusoidal responses of all the detectors to rotation of the
sparse wire grid. In each case, the differences between
the detector angles determined by the secondary method and the Moon 
are described by a standard deviation of $\approx 3^\circ$. 
However, we find a mean shift between  the Tau~A-derived and Moon-derived angles of $1\fdg7$.  To estimate the errors in the angles in light of this shift, we use an empirical approach: in Section~\ref{sec:gain_sys} we estimate the impact on the power spectra from using the Tau~A results instead of the Moon results, and find it to be small.

\begin{figure}[htbp]
\centering
\includegraphics[width=3.0in]{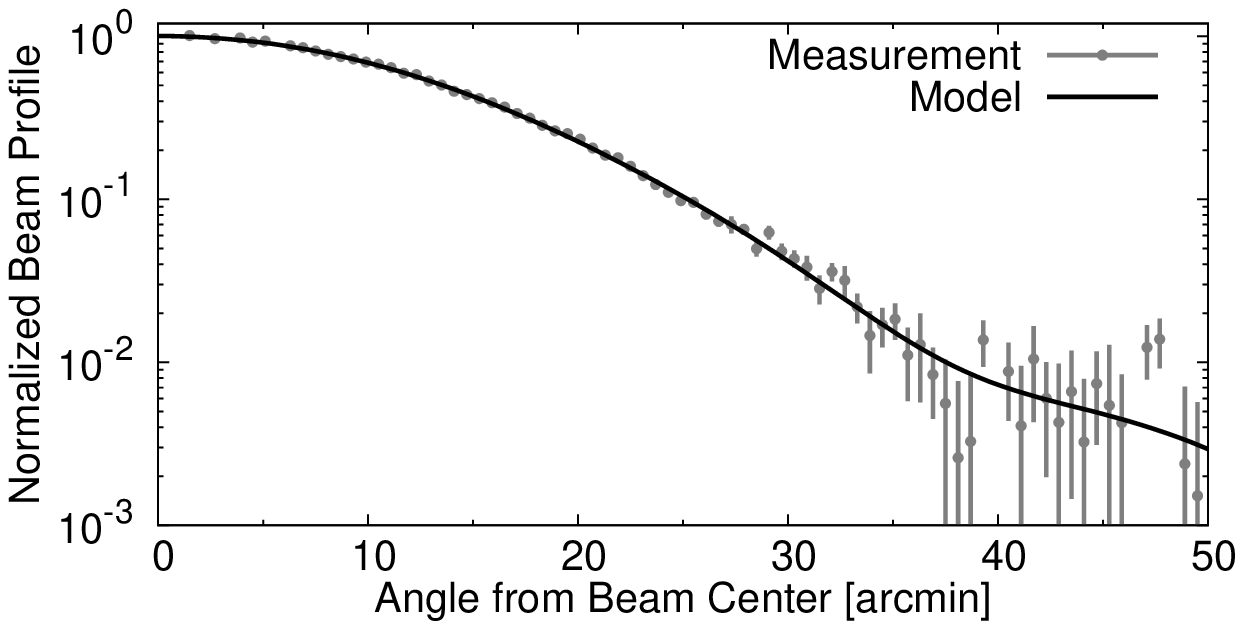}
\includegraphics[width=3.0in]{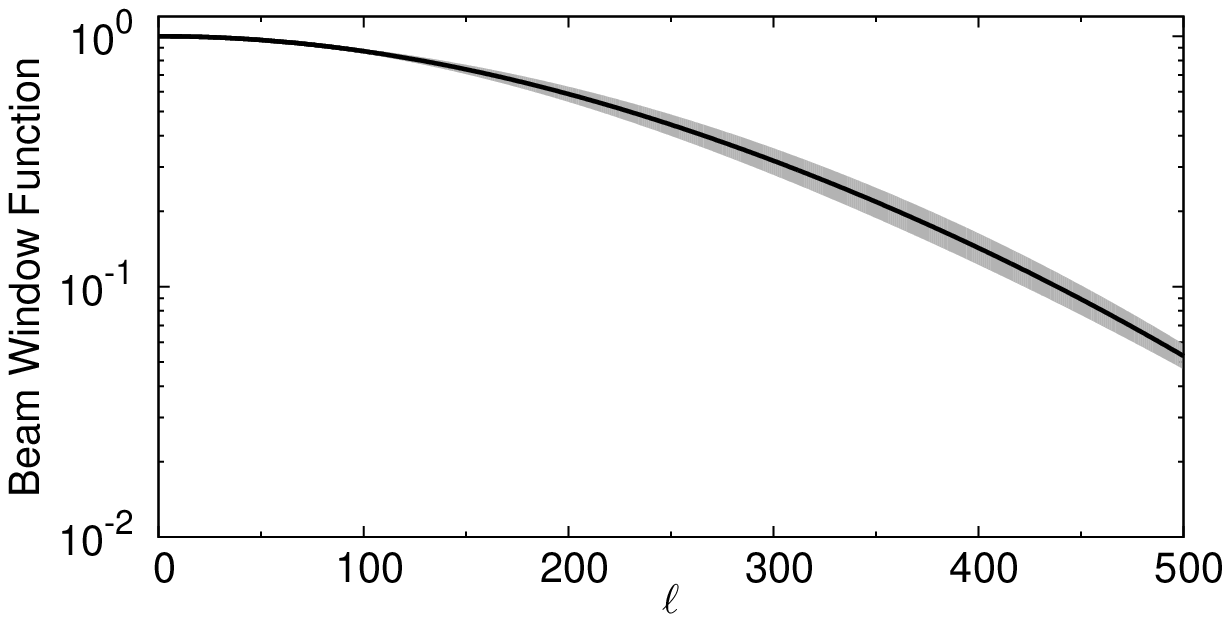}

 \vspace{-0.05in}

\caption{Top panel: Polarization beam profile from Tau~A observations with the central feed horn.  The data are overplotted with the expansion in Gauss-Hermite polynomials described in the text. 
Bottom panel: Beam window function with errors shown by the gray band.
         }
\label{fig:calibration_beam_and_window}
\end{figure}

 \vspace{0.5in}

\subsection{Beam Profile and Window Function}
\label{sec:beams}

The polarization and differential-temperature beams are obtained from maps created using the full data sets of Tau~A and Jupiter observations respectively, with square pixels of $1\farcm8$ on a side.  For polarization, this process produces the main and leakage beam maps simultaneously, with the latter describing the instrumental polarization. 
The average FWHM for the beams across the array is $27\farcm3$, 
measured with $0\farcm1$ precision
for the central feed horn and for the differential-temperature feed horns at the edge of the focal plane.  
The non-central--polarization-horn FWHMs  are measured less
frequently and thus are less precisely known, with an uncertainty of $1\farcm5$.
The beam elongation is typically small (1\%), and its effect is further reduced 
by the diurnal sky rotation and weekly deck rotations which result in a symmetrized
effective beam in the CMB maps.
We compute 1-dimensional symmetrized beam profiles, with a resolution of $0\farcm6$. 
These  profiles are modeled as a sum of six even Gauss-Hermite terms
\citep{monsalve:77412M}.
The main-beam solid angles are computed by integrating these models out to $54^\prime$ (roughly  $-28$\,dB),   yielding  $78.0\pm 0.4\,\mu\mbox{sr}$ for the differential-temperature horns and $74.3\,\pm\,0.7\,\mu\mbox{sr}$ for the central horn. 
An average gives $76\,\mu$sr for all horns in the array. 
We also examine alternative estimates such as integrating the raw
beam map instead of the analytical fit. We assign a systematic
uncertainty of 4\% based on the differences among these different
estimates.  The systematic error includes possible contributions from
sidelobes, which we constrain to $0.7\pm0.4\,\mu$sr with antenna range
measurements carried out before the observation season.

The window functions, encoding the effect of the finite resolution of the instrument on the power spectra, are computed
from the central-horn and the temperature-horn--profile models.  
The central-horn beam profile and window function are shown in 
Figure \ref{fig:calibration_beam_and_window}.
The uncertainty accounts for statistical error and differences between polarization and differential-temperature beams, as described in Section~\ref{sec:beam_sys}.

\subsection{Instrumental Polarization}
\label{sec:calibration_ip}

Instrumental imperfections can lead to a spurious polarization signal proportional to 
the unpolarized CMB temperature anisotropy.  We call this the   I to Q (or U) leakage term.   In our instrument,  a fraction of the power input on one port of the correlation module is reflected because of a bandpass mismatch to  the septum polarizer, and a fraction of the reflected power re-enters the other port.  The dominant monopole term comes from this effect. We measure the monopole term from the polarimeter responses to temperature changes, using sky dips; Moon, Tau~A, and Galactic signals; as well as variations from the weather.  The average magnitude is 1.0\%  (0.2\%) for the Q (U) diodes.  Note that the discrepancy  in the Q and U averages was predicted from measurements of the properties of the septum polarizers and confirmed in the field.  
We do not correct for this effect but assign systematic errors as
described in Section~\ref{sec:systematics_ip}.

\section{Data Analysis Procedure}
\label{sec:analysis}

QUIET employs two independent analysis pipelines to derive CMB power spectra. 
We present the methods used for analysis in each pipeline, including data selection, 
filtering, map making, and power-spectra estimation.

Pipeline A is based on the pseudo-\cl analysis framework, first
described by \cite{Hivon2002}, which is used by numerous
experiments \citep{Netterfield2002, Brown:2009uy, Chiang:2010,
  larson:2010, Lueker:2009rx}.  This pipeline made all analysis choices
   in accordance with a strict (blind) analysis validation policy described in Section~\ref{sec:nulls}.
An advantage
of the pseudo-\cl framework is computational efficiency, which is
critical for completing the more than 30 iterations of the null-test
suite. For the same reason, this pipeline is used for the systematic-error evaluations found in Section~\ref{sec:systematics}.  Pseudo-\cl analysis 
also enables us to perform cross correlation, making the resultant power spectra immune to possible misestimation of noise bias.

Pipeline B implements a maximum-likelihood framework
\citep[e.g.,][]{Tegmark1997,Bond1998}, 
which has a long history of use by CMB experiments
\citep[e.g.,][]{Mauskopf2000,
Page2007, Wu2007, bischoff:2008}. This framework yields minimum-variance estimates of the power spectra, naturally accounts for E/B
mixing, and directly provides the exact CMB likelihood
required for estimation of cosmological parameters, without the use of
analytical approximations. In addition to power spectra, it produces 
unbiased maps with full noise-covariance matrices, useful for comparisons with other experiments. 
On the other hand, this approach is also computationally
more expensive than the pseudo-\cl framework, and a reduced set
of null tests is therefore used to evaluate data consistency.

The processing of the time-ordered data (TOD) and the methodology used
for data selection are treated in Sections \ref{subsec:TOD} and 
\ref{subsec:cuts}, respectively. Brief descriptions of the pseudo-\cl 
and maximum-likelihood techniques are found in 
Section~\ref{sec:analysis_methods}.  TOD processing, data selection, and analysis for temperature-sensitive modules are discussed in Section~\ref{subsec:TT}.

 \vspace{0.5in}

\subsection{Time-Ordered--Data Processing}
\label{subsec:TOD}

To prepare the TOD for map making, we execute three steps: pre-processing, noise modeling, and filtering.   Of these steps, 
only the filtering is significantly different between the two pipelines.

\subsubsection{Pre-processing}
\label{sec:data:tod:preprocess}

The first data-processing step is to correct for a small
non-linearity that was discovered in the analog-to-digital converter (ADC) 
system. 
The non-linearities occur every 1024 bits; roughly 14\% of the data are affected.
Systematic uncertainty 
from this effect is estimated in Section~\ref{sec:systematics_other}.
Next, the receiver data are synchronized with the telescope pointing.
The double-demodulation step, described in
Section~\ref{sec:instrument}, is applied, reducing the sample rate from 100\,Hz
to 50\,Hz. A model of the detectors' polarized responsivities
converts the data from ADC counts into  thermodynamic temperature. The two pipelines
use different responsivity models.  Pipeline A applies a constant responsivity 
throughout each CES, addressing possible variability within a CES as part of the 
systematic error (Section~\ref{sec:systematics});  pipeline B updates responsivities on  2-minute  
timescales  \citep{dumoulin:77412N}.

\subsubsection{Noise Model}
\label{sec:data:tod:noise}

After pre-processing, the time streams for each detector diode in each CES are Fourier-transformed and their noise power spectra are fit to a
 model\footnote{At the level of a single CES, the TOD of each 
detector diode are dominated by noise; the contribution of the CMB is negligible.} with three parameters:
the amplitude of white noise, the $1/f$ knee frequency, and the
power-law slope of the $1/f$ noise.  We also compute the white-noise correlations 
among detector diodes in the same module:  the most important are between the two Q or the two U detector diodes (with an average coefficient of 0.22). A small fraction of the noise
spectra contain features not accounted for in the noise model: beam
sidelobes (see Section~\ref{sec:instrument}) scanning across features on the
ground create a narrow spike at the scan frequency; slowly-changing
weather patterns during a CES create a broader peak also at the scan
frequency; and there are some narrow spikes at high ($\gtrsim 6$\,Hz)
frequencies. To prevent these features from biasing the noise
model, the fit excludes a region around the scan frequency as well as
frequencies above 4.6\,Hz.  
In addition to the noise-model parameters, several statistics quantifying the agreement between the data and noise model are also  
 used for data selection as described in Section~\ref{subsec:cuts}.

\subsubsection{Filtering}
\label{sec:data:tod:filter}

 In pipeline A, three filters are applied. These
were chosen from the
results of many runs of the null-test suite (see Section~\ref{sec:nulls}).
First, to remove the high-frequency narrow spikes, we apply a low-pass
filter that cuts signals off sharply above 4.6\,Hz\footnote{For
QUIET's beam size and scanning speed a low-pass filter of 4.5--4.6\,Hz
results in a minimal loss of sensitivity to the CMB.}. Second, to
suppress contamination from atmospheric fluctuations and detector
$1/f$ noise, we subtract a linear function from each telescope
half scan (left-going or right-going) removing modes below twice the
scan frequency\footnote{Typical scan frequencies range from 45\,mHz to
100\,mHz.}.  The third filter, designed to eliminate signal from ground emission, removes any azimuthal structure that
remains after summing over all half scans in the CES. 

In pipeline B, an apodized bandpass filter is used that accepts modes
from 2.5 times the scan frequency to 4.5\,Hz; the highpass component
of this filter is designed to suppress scan-synchronous
contamination. Further, a time-independent ground-emission model is subtracted.
The model of ground emission is generated by building low-resolution and high--signal-to-noise maps in horizon coordinates from the full-season data
for each deck angle and module, using large ($55'$) pixels. Only
features that are stable in time, azimuth, elevation, and deck angle
contribute to this model. The amplitude of the ground correction is
$\lesssim1\,\mu\textrm{K}$.

\subsection{Data Selection}
\label{subsec:cuts}

The fundamental unit of data used for analysis is the
double-demodulated output of one detector diode for a single CES, referred to
as a ``CES-diode.'' Selecting only those CES-diodes that correspond to
good detector performance and observing conditions is a critical
aspect of the data analysis. The data-selection criteria began with a
nominal set of cuts and evolved into several distinct configurations,
as many as 33 in the case of pipeline A.
For each configuration, analysis validation (see Section~\ref{sec:nulls}) was performed yielding statistics quantifying the lack of contamination in the data set. The final data set was chosen when these statistics showed negligible contamination and were little affected by changes to the cuts.

Cut efficiencies, defined as the fractions of CES-diodes accepted for
the analysis, are given for both pipelines in Table
\ref{tbl:dataselection}. While each pipeline applies its own cuts
uniformly to all four patches, the efficiencies among patches are non-uniform
because of differences in weather quality.
 Over the course of the eight month observing
season, patch CMB-1 is primarily visible at night, when the
atmosphere tends to be more stable; patch CMB-3 is mostly observed during
the day.

The first step of the data selection is simply to remove known bad  
data: data from six non-functional detector diodes, data during periods of  mount malfunctions, and CESes lasting less than 1000\,s.
Further, we cut individual CES-diodes that show deviation from the expected 
linear relationship between the demodulated and TP signals.
This cut removes data with poor thermal regulation of the electronics or cryostat, or residual ADC non-linearity.

The beam sidelobes, described in Section~\ref{sec:instrument}, introduce
contamination to the data if the telescope scanning motion causes them
to pass over the ground or the Sun. Ground
pickup is dealt with 
by filtering as described in Section~\ref{sec:data:tod:filter}.  The less frequent cases of
Sun contamination are handled by cutting those CES-diodes for which the Sun's
position
overlaps with the measured sidelobe regions for each diode.

Additional cuts are specific to each pipeline.  Pipeline A removes
data taken during bad weather using a statistic calculated from
fluctuations of the TP data during 10-s periods, averaged across
the array.  This cut removes entire CESes.
Several more cuts remove individual CES-diodes.  While these
additional cuts are derived from the noise modeling statistics, they
also target residual bad weather. During such marginal weather
conditions only some channels need to be cut,
since the sensitivity for a given detector diode to atmospheric
fluctuations depends on its level of instrumental polarization. Next, we reject CES-diodes with poor agreement between the
filtered data and the noise model in three frequency ranges: a narrow
range (only 40 Fourier modes) about the scan frequency, from twice
the scan frequency to 1\,Hz, and from 1\,Hz to 4.6\,Hz. We also cut
CES-diodes that have higher than usual $1/f$ knee frequencies, or large variations during the CES in the azimuthal slopes of the double-demodulated time streams;  both these cuts help eliminate bad weather periods.  
Finally, we also remove any CES-diodes with
an outlier greater than $6\,\sigma$ in the time domain on three
timescales (20\,ms, 100\,ms, and 1\,s).

For pipeline B, the weather cut rejects CESes based on a statistic
computed from fluctuations of the double-demodulated signals from the
polarization modules on 10-s and 30-s timescales.  Three cuts are
applied to remove individual CES-diodes.  The first is a cut on the
$1/f$ knee frequency, similar to that of pipeline A. Second, a cut is
made on the noise model $\chi^2$ in the frequency range passed
by the filter, and third, we reject CES-diodes having a large
$\chi^2$ in the azimuth-binned TOD.  This cut rejects data with
possible time variation in the ground signal. Finally, an entire CES
is removed if more than 40\% of its detectors have already been rejected.

\begin{table}
\begin{center}
\caption{Total Hours Observed and Data-Selection Efficiencies}
\label{tbl:dataselection}

 \vspace{-0.05in}

\begin{tabular}{lcccc}
\hline \hline
Patch & Total Hours & A \% & B \% & Common \%\\
\hline
CMB-1 & 905 & 81.7 & 84.3 & 76.7\\
CMB-2 & 703 & 67.3 & 70.0 & 61.2\\
CMB-3 & 837 & 56.0 & 61.4 & 51.4 \\
CMB-4 & 223 & 70.6 & 74.2 & 65.9\\
\hline
All Patches & 2668 & 69.4 & 72.9 & 64.2 \\ 
\hline
\end{tabular}

 \vspace{-0.1in}

\end{center}
\tablecomments{Selection efficiencies for each pipeline.  ``Common'' gives the efficiencies if both sets of cuts were applied.}
\end{table}

\subsection{Map Making and Power-Spectra Estimation}
\label{sec:analysis_methods}

   After filtering, the TOD for all diodes are combined to produce $Q$ and $U$ maps for each of the QUIET patches. The maps use a HEALPix 
 $N_{\textrm{side}} = 256$ pixelization \citep{Gorski:2004by}.   This section describes the map making and power-spectra
estimation from the maps for each of the pipelines.  

\subsubsection{Pipeline-A Map Making}
\label{subsec:pcl}

Polarization maps ($Q$ and $U$) are made by summing samples into each pixel weighted by their inverse variance, calculated from the white-noise amplitudes. 
The full covariance matrix is not calculated. Two polarized sources, Centaurus A and Pictor A, are visible in the maps and are removed using circular top-hat masks with radii of $2\degr$ and $1\degr$, respectively.

Separate maps are made for each range of telescope azimuth and deck-angle orientations. The coordinates are binned such that there are 10 divisions in azimuth\footnote{The azimuth divisions are the same for all patches, which means that not all divisions are populated for patches CMB-3 and CMB-4.} and six distinct ranges of deck-angle orientation. Making separate maps for different telescope pointings enables the cross correlation described in the next section. 

\subsubsection{Power-Spectra Estimation in Pipeline A}
\label{subsec:pcl_ps}
 The MASTER (Monte Carlo Apodized Spherical Transform
Estimator) method is used in pipeline A
\citep{Hivon2002,Hansen:2002iha}; it is based on a pseudo-$C_\ell$
technique and takes account
of effects induced by the data processing using Monte Carlo (MC)
simulations.  The pseudo-$C_\ell$ method allows estimation of the
underlying
$C_\ell$ using spherical-harmonics transformations when the observations do
not cover the full sky uniformly \citep{Wandelt2001}.
The pseudo-$C_\ell$
spectrum, designated by
$\tilde{C}_\ell$, is related to the true spectrum $C_\ell$ by:
\begin{equation}
  \langle \tilde{C}_\ell \rangle = \sum_{\ell^\prime} M_{\ell \ell^\prime} F_{\ell^\prime} B_{\ell^\prime}^2 \langle C_{\ell^\prime} \rangle.
  \label{eq:pseudo_cl_temperature}
\end{equation}
There is no term corresponding to noise bias, which would arise if we
did not employ a cross-correlation technique.
Here $B_\ell$ is the beam window function, described
in Section~\ref{sec:beams}, and $M_{\ell \ell^\prime}$ is a mode-mode--coupling
kernel describing the effect of observing only a small fraction of the
sky with non-uniform coverage.  It is calculable from the pixel weights,
which are chosen to maximize the signal-to-noise ratio \citep{FKP}.
We bin in $\ell$ and recover $C_\ell$ in nine band powers, $C_b$, 
and $F_\ell$ is the transfer function (displayed in Section \ref {sec:results})  due to filtering of the data;
its binned estimate, $F_b$, is found by processing noiseless CMB simulations
through pipeline A and used to obtain $C_b$.  For the polarization power spectra, equation
(\ref{eq:pseudo_cl_temperature}) is generalized for the case
where $\tilde{C}_\ell$ contains both $\tilde{C}_\ell^{EE}$ and
$\tilde{C}_\ell^{BB}$.

In the power-spectra estimates, we include only the cross correlations
among pointing-division maps, excluding the auto correlations.
Because the noise is uncorrelated for different pointing 
divisions, the cross-correlation technique allows us to eliminate
the noise-bias term and thus the possible residual bias due to its
misestimate.  Cross correlation between different pointing divisions also 
suppresses possible effects  of ground contamination and/or time-varying effects.
Dropping the auto correlations creates only a small increase in the statistical 
errors ($\approx 3$\%) on the power spectra.

The errors estimated for the pipeline-A power spectra are frequentist
two-sided 68\% confidence intervals. A likelihood function used to
compute the confidence intervals is modeled following
\cite{Hamimeche:2008ai} and calibrated using the MC simulation ensemble of
more than 2000 realizations with and without CMB signal.
We also use the likelihood function to put constraints on $r$ and calculate the consistency to $\Lambda$CDM.

The partial sky coverage of QUIET generates a small amount of E/B mixing 
 \citep{Challinor:2004pr}, which contributes an additional variance to the BB power spectrum.  We incorporate it as part of the statistical
error.  This mixing can be corrected \citep{Smith:2006vq} in future
experiments where the effect is not negligible compared to
instrumental noise.

\subsubsection{Pipeline-B Map Making}
\label{subsec:analysis_ml}

In pipeline B, the pixel-space sky map
$\hat{\mathbf{m}}$ ($N_{\textrm{side}} = 256$) is given by
\begin{equation}
\hat{\mathbf{m}} = 
\left(\mathbf{P}^{T}\mathbf{N}^{-1}\mathbf{F}\mathbf{P}\right)^{-1} 
\mathbf{P}^{T}\mathbf{N}^{-1} \mathbf{F} \mathbf{d},
\label{eq:mlmapmaking}
\end{equation}
where $\mathbf{P}$ is the pointing matrix, $\mathbf{N}$ is the TOD--noise-covariance matrix, $\mathbf{F}$ corresponds to the apodized bandpass filter
discussed in Section~\ref{sec:data:tod:filter}, and $\mathbf{d}$ denotes the
TOD. This map is unbiased, and for the case
$\mathbf{F}=\mathbf{1}$ it is additionally the maximum-likelihood map,
maximizing
\begin{equation}
\mathcal{L}({\mathbf{m}}|\mathbf{d})
=\mathrm{e}^{-\frac{1}{2}\left(\mathbf{d}-\mathbf{Pm}\right)^T\mathbf{N}^{-1}\left(\mathbf{d}-\mathbf{Pm}\right)}.
\end{equation}
The corresponding map--noise-covariance
matrix \citep[e.g.,][]{Tegmark1997,keskitalo:2010} is 
\begin{equation}
\mathbf{N_{\hat{m}}} =
\left(\mathbf{P}^{T}\mathbf{N}^{-1}\mathbf{F}\mathbf{P}\right)^{-1} 
\left(\mathbf{P}^{T}\mathbf{F}^{T}\mathbf{N}^{-1}\mathbf{F}\mathbf{P}\right) 
\left(\mathbf{P}^{T}\mathbf{N}^{-1}\mathbf{F}\mathbf{P}\right)^{-1} . 
\label{eq:mlcovmat}
\end{equation}
Note that one often encounters the simplified expression
$\mathbf{N_{\hat{m}}} =
\left(\mathbf{P}^{T}\mathbf{N}^{-1}\mathbf{F}\mathbf{P}\right)^{-1}$
in the literature. This corresponds effectively to assuming that
$\mathbf{F} = \mathbf{F}^2$ in the Fourier domain, and is strictly
valid for top-hat--filter functions only. For our filters, we find
that the simplified expression biases the map-domain $\chi^2
 (\equiv \hat{\mathbf{n}}^T \mathbf{N_{\hat{m}}^{-1}}\hat{\mathbf{n}}$, where
$\hat{\mathbf{n}}$ is a noise-only map) by $\approx 3\,\sigma$,
and we therefore use the full expression, which does lead to an
unbiased $\chi^2$.

Equations (\ref{eq:mlmapmaking}--\ref{eq:mlcovmat}) apply to both
polarization and temperature analysis. The only significant difference
lies in the definition of the pointing matrix, $\mathbf{P}$. For
polarization, $\mathbf{P}$ encodes the detector orientation, while for
temperature it contains two entries per time sample, $+1$ and $-1$,
corresponding to the two horns in the differential-temperature assembly.

After map making, the maps are post-processed by removing unwanted
pixels (i.e., compact sources and low--signal-to-noise edge pixels).  All
54 compact sources in the 7-year \textit{WMAP} point source
catalog \citep{gold:2010} present in our four patches are masked out,
for a total of 4\% of the observed area.  We also marginalize over
large-scale and unobserved modes by projecting out all modes with
$\ell \le 5$ ($\ell \le 25$ for temperature) from the noise-covariance
matrix using the Woodbury formula, assigning infinite variance to
these modes.

\subsubsection{Power-Spectra Estimation in Pipeline B }
\label{subsec:ml_ps}
Given the unbiased map estimate, $\hat{\mathbf{m}}$, and
its noise-covariance matrix, $\mathbf{N_{\hat{m}}}$, we estimate the
binned CMB power spectra, $C_b$, using the Newton--Raphson
optimization algorithm described by \cite{Bond1998}, generalized to
include polarization. In this algorithm one iterates towards the
maximum-likelihood spectra by means of a local quadratic
approximation to the full likelihood. The iteration scheme in its
simplest form is
\begin{equation}
\delta C_b = \frac{1}{2} \sum_{b'}
\mathcal{F}^{-1}_{bb'} \textrm{Tr}\left[(\hat{\mathbf{m}}\hat{\mathbf{m}}^T
- \mathbf{C})(\mathbf{C}^{-1}\mathbf{C}_{,b'}\mathbf{C}^{-1})\right],
\end{equation}
where $b$ denotes a multipole bin, $\mathbf{C}$ is the
signal-plus-noise pixel-space covariance matrix, and $\mathbf{C}_{,b}$
is the derivative of $\mathbf{C}$ with respect to $C_b$. The signal
component of $\mathbf{C}$ is computed from the binned power spectra,
$C_b$, and the noise component is based on the noise model described
in Section \ref{sec:data:tod:noise}, including diode-diode
correlations. Finally, 
\begin{equation}
\mathcal{F}_{bb'} = \frac{1}{2} \textrm{Tr}(\mathbf{C}^{-1}\mathbf{C}_{,b}\mathbf{C}^{-1}\mathbf{C}_{,b'})
\end{equation}
is the Fisher matrix. Additionally, we introduce a step length
multiplier, $\alpha$, such that the actual step taken at iteration $i$
is $\alpha \, \delta C_b$, where $0 < \alpha \le 1$ guarantees that
$\mathbf{C}$ is positive definite. We adopt the diagonal elements of
the Fisher matrix as the uncertainties on the band powers.

We start the Newton--Raphson search at $\cl = 0$, and iterate until
the change in the likelihood value is lower than 0.01 times the number
of free parameters, corresponding roughly to a 0.01-$\sigma$
uncertainty in the position of the multivariate peak. Typically we
find that 3 to 10 iterations are required for convergence.

Estimation of cosmological parameters, $\theta$, is done by
brute-force grid evaluation of the pixel-space likelihood,
\begin{equation}
\mathcal{L}(\theta) \propto \frac{-\frac{1}{2}\mathbf{d}^{T}\mathbf{C}^{-1}(\theta)\mathbf{d}}{\sqrt{|\mathbf{C}(\theta)|}}.
\end{equation}
Here $\mathbf{C}(\theta)$ is the covariance matrix evaluated with a
smooth spectrum, $C_{\ell}$, parametrized by $\theta$. In this paper,
we only consider 1-dimensional likelihoods with a parametrized
spectrum of the form $C_{\ell} = a\, C_{\ell}^{\textrm{fid}}$, $a$
being a scale factor and $C_{\ell}^{\textrm{fid}}$ a
reference spectrum; the computational expense is therefore not a
limiting factor.  Two different cases are considered, with $a$ being
either the tensor-to-scalar ratio, $r$, or the amplitude of the EE
spectrum, $q$, relative to the $\Lambda$CDM model.

\subsection{Temperature Data Selection and Analysis}
\label{subsec:TT}

As described in Section~\ref{sec:instrument}, we dedicate one pair of modules to differential-temperature measurements.  While these modules are useful for calibration purposes, when combined with our polarization data they also enable us to make self-contained measurements of the TE and TB power spectra.  

For temperature, both pipelines adopt the pipeline-A data-selection criteria used for polarization analysis (see Section~\ref{subsec:cuts}). 
  The temperature-sensitive modules, however, are far more susceptible to atmospheric 
contamination than the polarization modules.  Thus, these cuts result in reduced efficiencies: 12.4\%, 6.9\%, and 6.8\% for patches CMB-1, CMB-2, and CMB-3, respectively\footnote{Patch CMB-4 is excluded due to low data-selection efficiency and a lack of sufficient crosslinking.}.
More tailoring of the cuts for these modules would improve efficiencies.

In pipeline A, the analysis proceeds as described in Sections \ref{sec:data:tod:filter}, \ref{subsec:pcl}, and \ref{subsec:pcl_ps} except for two aspects.  First, in the TOD processing a
second-order polynomial is fit and removed from each telescope half scan instead of a linear
function.  This suppresses the increased contamination from atmospheric fluctuations in the temperature data.  Second, we employ an iterative map maker based on the algorithm described by \citet{wright:1996}.  
Map making for differential receivers requires that each pixel is measured at multiple array pointings or crosslinked.  
In order to improve crosslinking we divide the temperature data into only four maps by azimuth and deck angle, rather than the 60 divisions used for polarization analysis.
To calculate TE and TB power spectra, polarization maps are made for these four divisions, plus one additional map that contains all polarization data with pointings not represented in the temperature data.

For pipeline B the algorithms for making temperature maps and
estimating power spectra are identical to the polarization case, as
described in Sections \ref{subsec:analysis_ml} and \ref{subsec:ml_ps}.

\section{Analysis Validation}
\label{sec:nulls}

The QUIET data analysis follows a policy of not looking at the power spectra until the analysis is validated using a set of predefined tests for possible systematic effects\footnote{Some systematic effects, such as a uniform responsivity-calibration error, cannot be detected by these techniques, and are addressed in Section~\ref{sec:systematics}.}.
The validation tests consist of a suite of null tests, comparisons across multiple analysis configurations, and consistency checks among power spectra from different CMB patches.
Data-selection criteria, filtering methods, and the division of data into maps for cross correlation in pipeline A are all evaluated based on the test results.  

Details of tests found in this section describe pipeline A. While the pipeline B analysis follows a similar program of null tests to verify the result, the increased computational requirements of the maximum-likelihood framework limit the number of tests that could be performed and require those tests to be run using lower-resolution maps than for the non-null analysis. The bulk of this section treats validation of the polarization analysis; at the end, we briefly describe the temperature analysis validation.

We conduct this validation in a blind-analysis framework to reduce
experimenter bias, the influence of the experimenter's knowledge of
prior results and theoretical predictions on the result (power spectra).
Blind analysis, making the analysis choices without knowing the result,
is a standard technique for minimizing this bias~\cite[]{2005ARNPS..55..141K}.
In our blind analysis framework, we finalize all aspects of the data
analysis including calibration, data selection, and evaluation of the
systematic error. Only after the analysis is finalized and the following
validation tests pass do we examine the power spectra.

In a null test, the data are split into two subsets. Maps, $m_1$ and $m_2$, are made from each subset. The power spectra of the difference map, $m_{\textrm{diff}} \equiv ( m_1 - m_2 ) / 2$, are analyzed for consistency with the hypothesis of zero signal. 
The null suite consists of 42 tests\footnote{Only 41 null tests are performed for patch CMB-4; one test is dropped because there are no data in one of the subsets.}, each targeting a possible source of signal contamination or miscalibration. 
These are highly independent tests; the data divisions for different null tests are correlated at only 8.8\% on average. 
Nine tests divide the data by detector diode based on susceptibility to instrumental effects, such as instrumental polarization. 
Ten tests target effects that depend on the telescope pointing such as data taken at high or low elevation.
Five tests divide based on the proximity of the main or sidelobe beams to known sources such as the Sun and Moon. 
Eight tests target residual contamination in the TOD using statistics mentioned in Section~\ref{subsec:cuts}. 
Ten tests divide the data by environmental conditions such as ambient temperature or humidity. 

\begin{figure}[htbp]
  \centering
  \includegraphics[width=\linewidth]{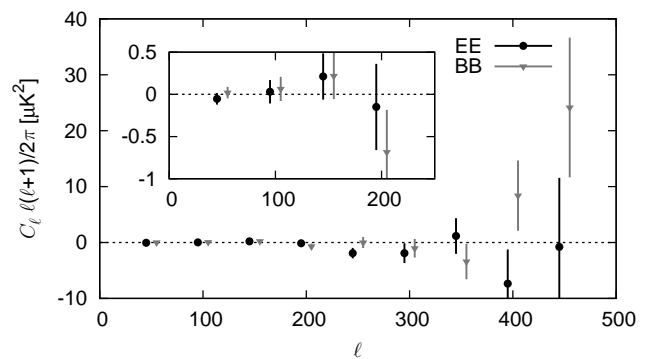}
  \caption{ \label{fig:nulls1} 
  EE and BB power spectra for the patch CMB-1 null test between Q and U detector diodes.  The inset shows the low-$\ell$ region in detail.}
\end{figure}

Each null test yields EE and BB power spectra in nine $\ell$ bins, calculated separately for each CMB patch. 
Figure \ref{fig:nulls1} shows the power spectra from one null test. 
Although the EB spectra are also calculated for each null test, they are assigned lesser significance since sources of spurious EB power will also result in the failure of  EE and BB null tests.
Combining all EE and BB points for all patches and null tests in the null suite yields a total of 3006 null-spectrum points. 
For each power-spectrum bin $b$, we calculate the statistic $\chi_\textrm{null} \equiv C^\textrm{null}_b / \sigma_{b}$, where $C^\textrm{null}_b$ is the null power and  $ \sigma_b$ is the standard deviation of $C^\textrm{null}_b$ in MC simulations. 
We evaluate both $\chi_\textrm{null}$ and its square; $\chi_\textrm{null}$ is sensitive to systematic biases in the null spectra while $\chi^2_\textrm{null}$ is more responsive to outliers.
We run MC simulations of the full null suite to take into account the small correlation among the null tests and the slight non-Gaussianity of the  $\chi_\textrm{null}$ distribution.  Non-Gaussianity is caused by the small number of modes at low $\ell$.

\begin{figure}[htbp]
  \begin{center}
    \includegraphics[width=\linewidth]{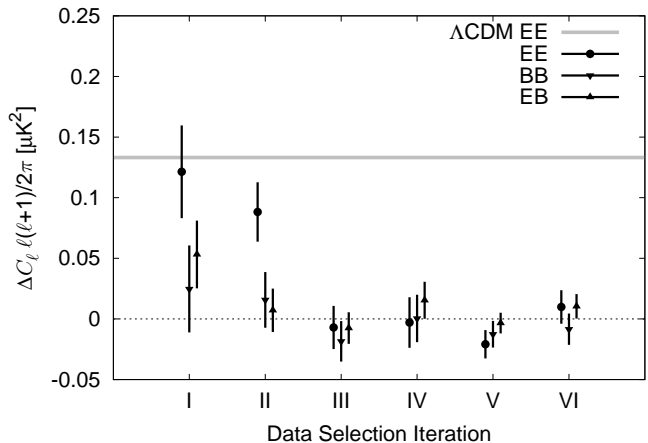}
  \end{center}

 \vspace{-0.15in}

  \caption{\label{fig:blind_config_comparison} 
  Power-spectra differences between the final data selection and six of the 32 earlier data-selection iterations, ordered by date.
  The lowest-$\ell$ bin of patch CMB-1 is shown.
  The error bars correspond to the expected fluctuations due to the differences in data selected, which are much smaller than the final statistical errors in this bin ($\approx0.10\,\mu\mbox{K}^2$ for BB).   Iterations that are closer to the final data selection have smaller errors.
  The expected EE power in this bin from the $\Lambda$CDM model is also shown for comparison.
}
\end{figure}

As we refine the data-selection criteria based on the results of the null suite,  we use a second test  to monitor changes in the non-null power spectra. 
Using a blind analysis framework, we compute the difference of the power spectra between any two iterations of the data selection without revealing the non-null spectra. Further, we randomize the sign of the difference to hide the  direction of the change; knowledge of the direction could allow experimenter bias (e.g. a preference for low BB power).
Figure \ref{fig:blind_config_comparison} shows the differences in the power spectra between the final configuration and several intermediate iterations of the data selection, starting with data sets that showed significant failures for the null-test suite.  Statistically significant differences indicate a change in the level of contamination in the selected data set.   Our data-selection criteria are finalized when further iterations only result in statistically expected fluctuations.  The sensitivity of this test is demonstrated by the fact that  the expected fluctuations are much less than the statistical error of the final result.

Finally, the non-null power spectra are compared among the four CMB patches. 
A $\chi^2$ statistic is computed from the deviation of each patch's non-null power spectra from the weighted average over all patches.  The total $\chi^2$ is compared to MC simulations to compute probabilities to exceed (PTE).

\begin{figure}[htbp]
  \begin{center}
    \includegraphics[width=\linewidth]{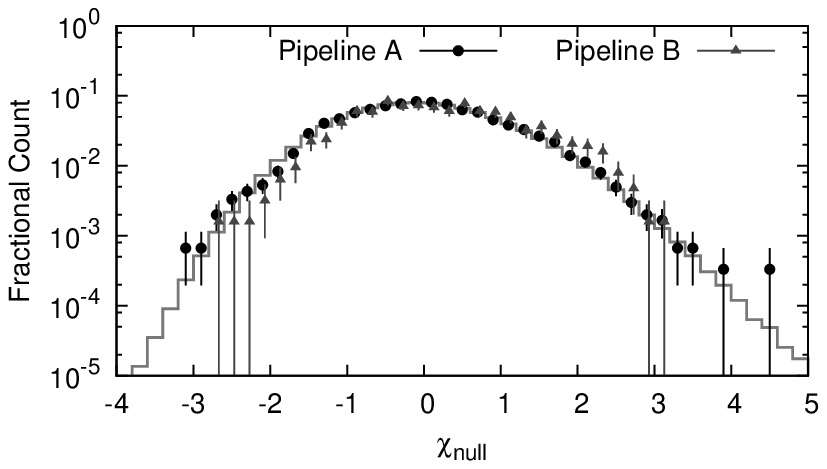}

 \vspace{-0.1in}

    \includegraphics[width=\linewidth]{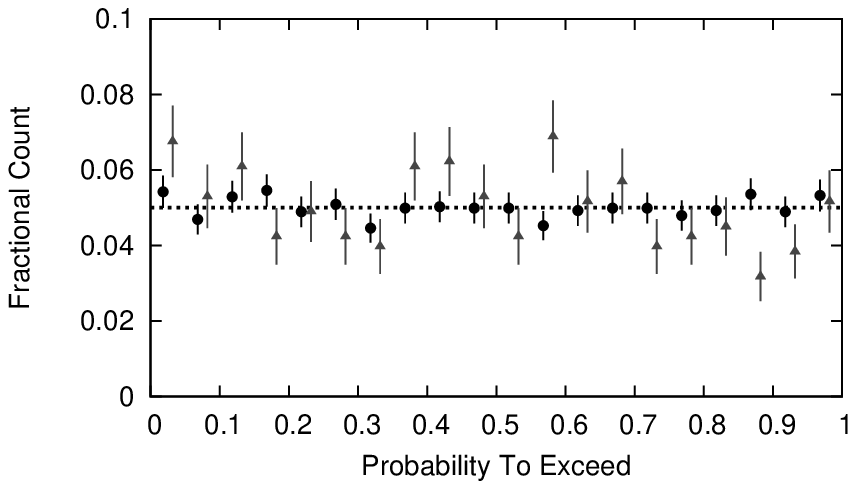}
  \end{center}

 \vspace{-0.15in}

  \caption{ \label{fig:nulls2} 
  Null-Suite Statistics.  The upper panel shows a histogram of the $\chi_{\textrm{null}}$ values for the pipeline-A null suite (circles), pipeline-B  null suite (triangles), and the average of 1024 MC realizations of the pipeline-A null suite (gray histogram). 
  Both data and MC distributions show similar non-Gaussianity in the $\chi_{\textrm{null}}$ statistic. The shift in $\chi_{\textrm{null}}$ seen for pipeline B, also seen in earlier iterations of pipeline A, is discussed in the text.
  The lower panel shows a histogram of PTEs calculated from the $\chi^2_{\textrm{null}}$ statistic (outliers from either side of the upper distribution manifest as low PTEs). 
  }
\end{figure}

When all aspects of the analysis are finalized, the last round of null tests and CMB patch comparisons validates the non-null--power-spectra results.
Figure \ref{fig:nulls2} shows the distributions of the $\chi_{\textrm{null}}$ statistic and of the PTEs corresponding to all $\chi^2_{\textrm{null}}$ values from the full null suite.   
In pipeline A, the distribution of $\chi_{\textrm{null}}$ is consistent with the expectation from MC simulations.  The mean of the $\chi_{\textrm{null}}$ distribution is $0.02\pm 0.02$; the mean of the MC-ensemble $\chi_{\textrm{null}}$ distribution is also consistent with zero.  
The distribution of the $\chi^2_\textrm{null}$  PTEs is uniform as expected.  
Table~\ref{tbl:null_chi2_by_patch} lists the PTEs for the sums of the $\chi^2_\textrm{null}$ statistic over all bins in each patch.
Examinations of various subsets of the null suite, such as EE or BB only, do not reveal any anomalies.  The EB null spectra do not indicate any failure either. Patch comparison PTEs are 0.16, 0.93, and 0.40 for EE, BB, and EB, respectively, demonstrating no statistically significant difference among the patches.

A similar, but smaller, null suite is run by pipeline B. Specifically,
21 null tests are made at a HEALPix resolution of
$N_{\textrm{side}}=128$.  The results obtained in these calculations
are summarized in the bottom panel of Figure \ref{fig:nulls2}, and
total PTEs for each patch are listed in
Table~\ref{tbl:null_chi2_by_patch}. As in pipeline A, no anomalous
values are found.

Finally, we make a comment on the usefulness of the
$\chi_{\textrm{null}}$ distribution (as opposed to the $\chi^2_{\textrm{null}}$
distribution) for identifying and quantifying potential
contaminants. During the blind stage of the analysis, a positive bias
in the $\chi_{\textrm{null}}$ distribution of 0.21 (0.19) was
identified using pipeline A (B) (corresponding to 21\% (19\%) of the
statistical errors). The number from pipeline A was obtained when
including auto correlations in its power-spectra estimator. When
excluding auto correlations, and cross-correlating maps made from data divided by time (day by day), the bias decreased to 0.10. Further detailed studies  lead to the division of data into maps based on the telescope pointing, as described in Section~\ref{sec:analysis_methods};  the result is an elimination of the observed bias.

The maximum-likelihood technique employed by pipeline B intrinsically
uses auto correlations, and a corresponding shift in the
$\chi_\textrm{null}$ distribution is seen in
Figure \ref{fig:nulls2}.  However, as will be seen in Section~\ref{sec:results}, the power spectra from the two pipelines are in excellent agreement, thereby confirming that any systematic bias coming from including auto correlations is well below the level of the statistical errors.  We close this section by mentioning that we
know of no other CMB experiment reporting an examination of the
$\chi_\textrm{null}$ distribution, which is sensitive to problems not
detected by examining the $\chi^2_\textrm{null}$ distribution only.

\begin{table}[t]
  \begin{center}
   \caption{\label{tbl:null_chi2_by_patch} Null Suite Probability To Exceed by Patch}
 
 \vspace{-0.05in} 

    \begin{tabular}{ccc}
      \hline
      \hline
      Patch & Pipeline A \% & Pipeline B \% \\
      \hline
      CMB-1 & 44 & 7 \\
      CMB-2 & 19 & 43 \\
      CMB-3 & 16 & 23 \\
      CMB-4 & 68 & 28 \\
      \hline
    \end{tabular}
  \end{center}

 \vspace{-0.1in}

\tablecomments{  PTEs  calculated from the sums of the $\chi^2_\textrm{null}$ statistics, for EE and BB spectra points, over the null tests for each patch.
  }

\end{table}

\subsection{Validation of the Temperature Analysis}
\label{sec:tt_blind_analysis}

A smaller number of null tests is used for the temperature analysis.
Several are not applicable 
 and others are discarded due to lack of data with sufficient crosslinking. 
Even so, we are able to run suites of 29, 27, and
23 TT null tests on patches CMB-1, CMB-2, and CMB-3, respectively.  We
calculate the sums of $\chi^2_\textrm{null}$ statistics, yielding PTEs
of 0.26 and 0.11 for patches CMB-1 and CMB-2, respectively.  No
significant outliers are found for these patches.  However, a
5-$\sigma$ outlier in a single test\footnote{This null test divides the data based on array pointing.}
is found in patch CMB-3, implying contamination in its temperature map.  CMB-3 is therefore excluded from further analysis.  We confirm consistency
between the patches CMB-1 and CMB-2 with a PTE of 0.26.

With no significant contamination in TT, EE, or BB spectra, one may be confident that the TE and TB spectra are similarly clean.
For confirmation, we calculate TE and TB null spectra for the five null tests that are common to the temperature and
polarization analyses.  
These yield PTEs of 0.61 and 0.82 for TE, and
0.16 and 0.55 for TB, for patches CMB-1 and CMB-2, respectively, with no significant outliers.
Patch consistency checks give PTEs of 0.48 for TE and 0.26 for TB.  Thus, the TE and TB power spectra, as well as the TT, pass all validation tests that are performed.

\section{Results}
\label{sec:results}

We report results from the first season of QUIET Q-band
observations: CMB power spectra, derived foreground estimates, and
constraints on the tensor-to-scalar ratio, $r$.

\subsection{Polarization Power Spectra}
\label{sec:pol_powspec}

\begin{figure}[!]
\label{fig:results}
\vspace{-0in}
\centering
\vspace{-0.0in}
\includegraphics[width=\linewidth]{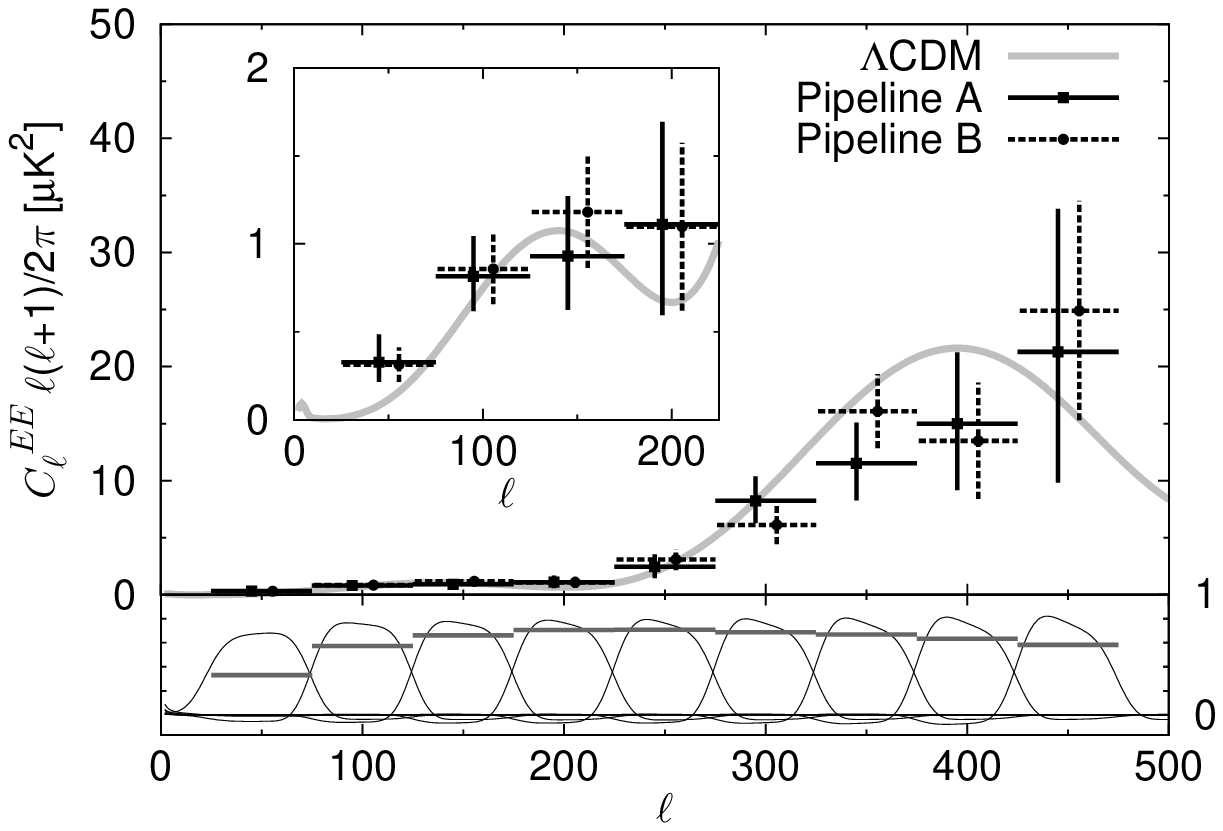}

 \vspace*{-0.1in}

\includegraphics[width=\linewidth]{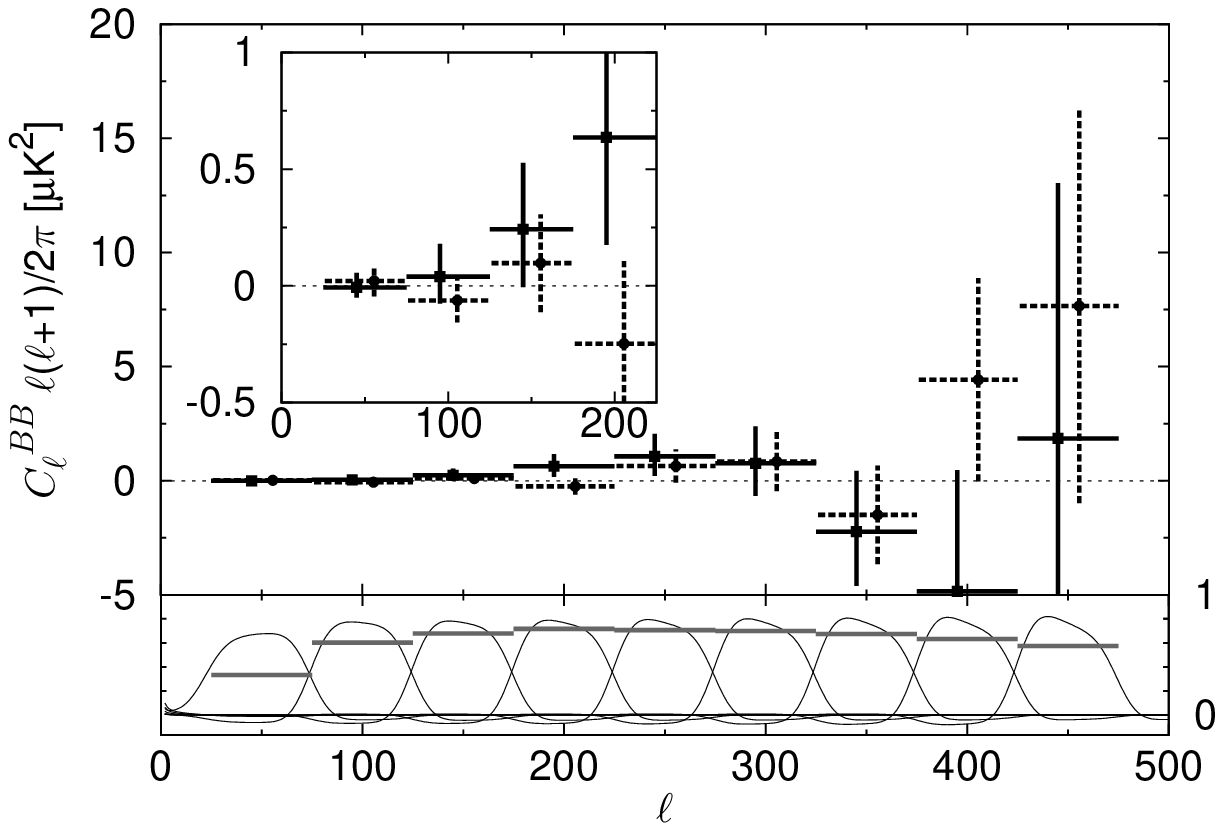}

 \vspace*{-0.1in}

\includegraphics[width=\linewidth]{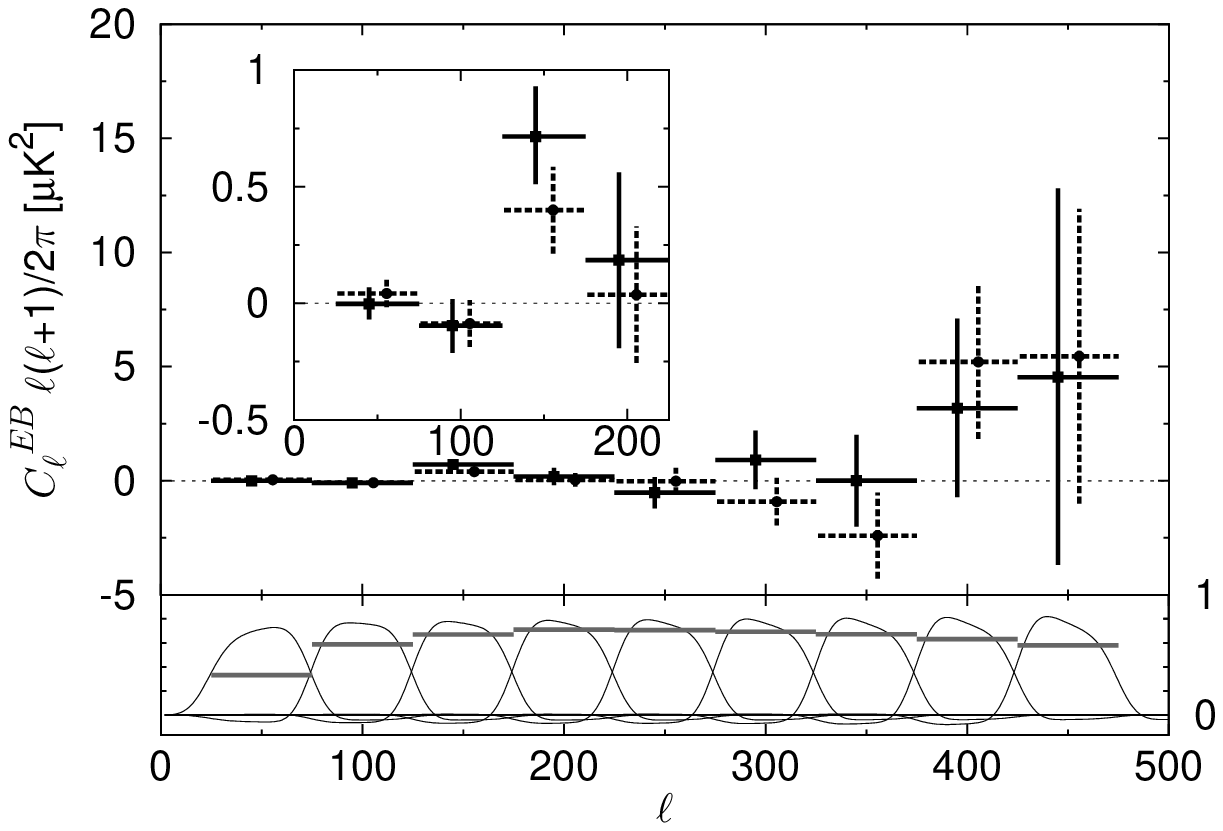}

 \vspace*{-0.1in}

\caption{ EE, BB, and EB power spectra from each  QUIET pipeline, all four patches combined.
 The insets show the low-$\ell$ region in
detail.  Window and transfer functions for each $\ell$ bin are shown
below the corresponding power spectra in black and gray,
respectively.
The window function combines the mode-mode--coupling kernel $M_{\ell \ell'}$ with the beam
($B_\ell$) and represents, in combination with the transfer function ($F_\ell$), the response in each
band to the true $C_\ell$ spectrum.
The EE point in the lowest-$\ell$ bin includes
foreground contamination from patch CMB-1.   For this display, pipeline A
shows frequentist 68\% confidence intervals while pipeline B uses the
diagonal elements of the Fisher matrix;  the difference is most pronounced in the lowest-$\ell$ bin where the likelihood
is the most non-Gaussian. }

\label{fig:result_summary}
\end{figure}

The CMB power spectra are reported in nine equally-spaced bands with 
$\Delta\ell=50$, beginning at  $\ell_{\textrm{min}}=25$.  
Given the patch size, modes with $\ell < \ell_{\textrm{min}}$
cannot be measured reliably.   The
correlation between neighboring bins is typically $-0.1$;  it becomes negligible for bins further
apart.

The EE, BB, and EB polarization power spectra estimated by both
pipelines are shown in Figure \ref{fig:result_summary}. The agreement
between the results obtained by the two pipelines is excellent, and
both are consistent with the $\Lambda$CDM concordance cosmology.  Our
findings and conclusions are thus fully supported by both pipelines.  Only the statistical
uncertainties are shown here; we treat systematic errors
in Section~\ref{sec:systematics}.  Because the systematic error analysis was only done for pipeline A, we adopt its power-spectra results (tabulated in Table \ref{tab:results}) as the official QUIET results.   

\begin{figure}[t]
\vspace{-0in}
\centering
\includegraphics[width=\linewidth]{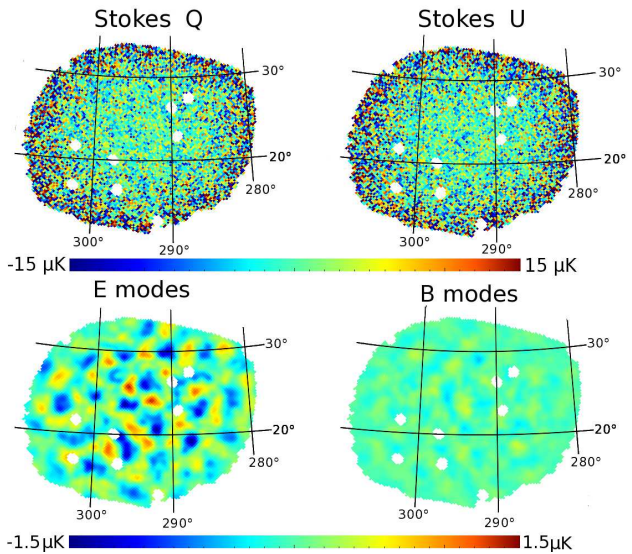}
\caption{Maps of patch CMB-1 in Galactic coordinates.    
The top row shows our polarization maps with compact sources masked (white disks).
The bottom row shows E and B modes decomposed using a generalized Wiener filter
technique, implemented through Gibbs sampling where the signal term
 of the Wiener filter is marginalized over the power spectra constrained
 by the data of this patch themselves \citep{eriksen:2004,
larson:2007}.  The maps include only modes for $\ell \ge 76$ and smoothed to
$1\degr$ FWHM; lower multipoles are removed due to a significant
foreground contribution. Note the clear difference in amplitude: the E modes show a high--signal-to-noise cosmological signal while the B modes are consistent with noise.
 Maps for the other patches are available online.
}
\label{fig:maps}
\end{figure}

\begin{figure*}[htbp]
\centering
\includegraphics[width=\linewidth]{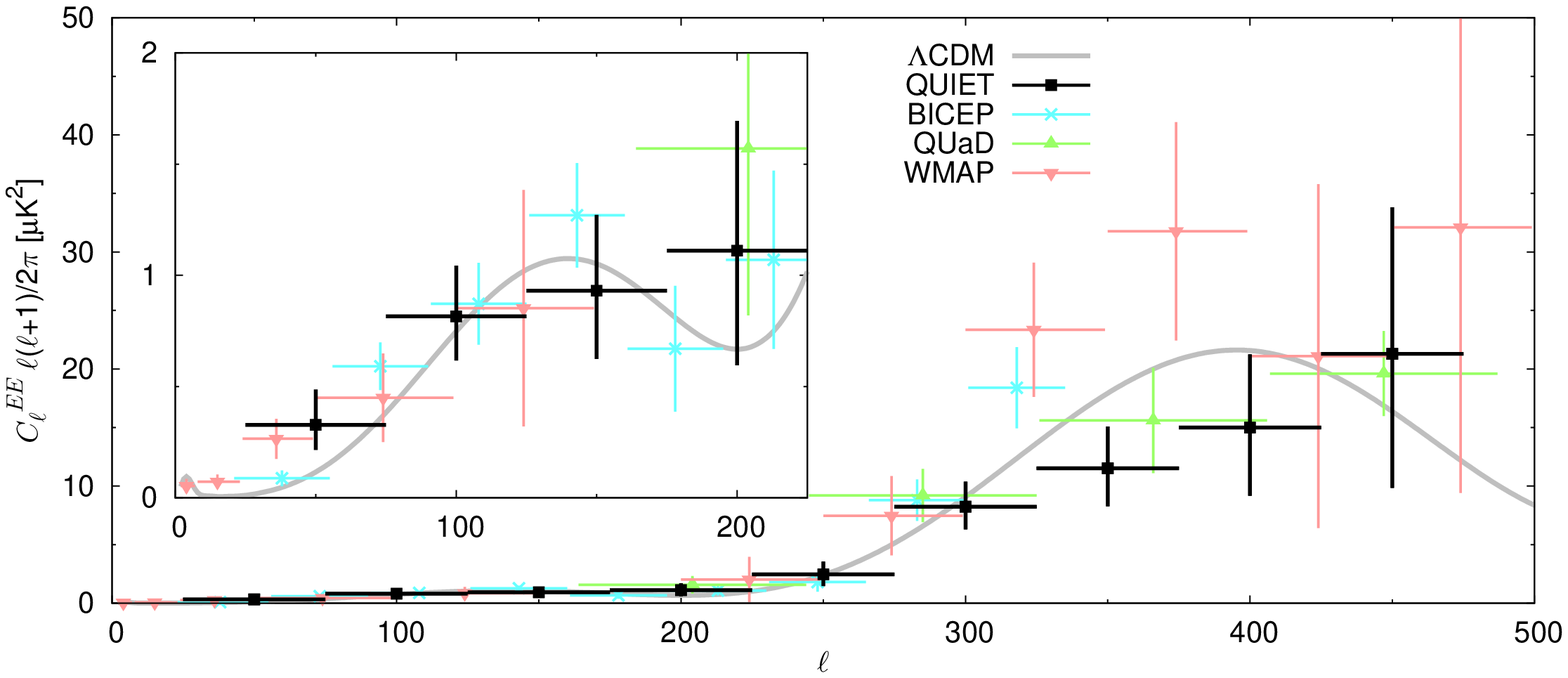}

 \vspace{-0.1in}

\includegraphics[width=\linewidth]{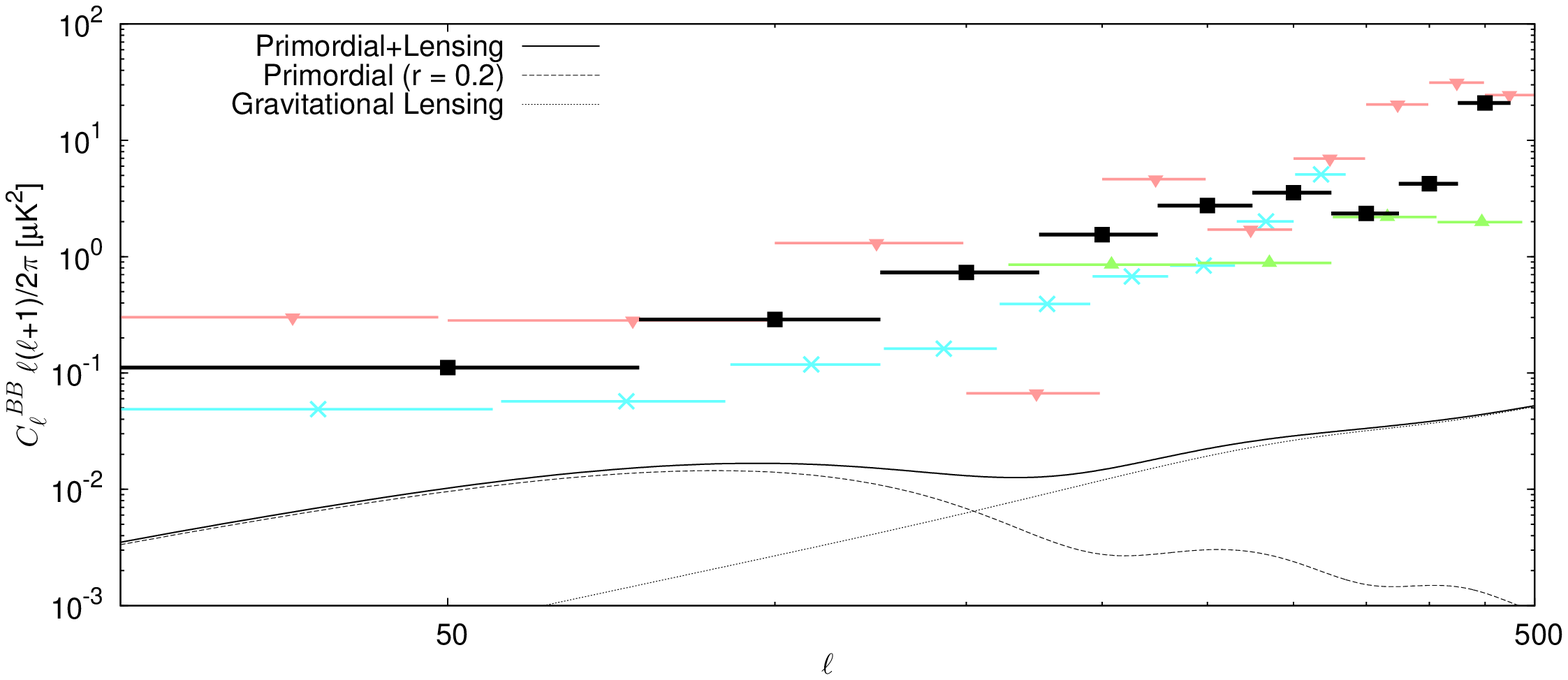}

 \vspace{-0.1in}

\caption{The top panel shows EE results with 68\% C.L. error bars; the bottom panel shows BB 95\% C.L. upper limits.  For comparison, we also plot results from previous experiments \citep{Brown:2009uy, Chiang:2010, larson:2010} and the $\Lambda$CDM model (the value $r=0.2$ is currently the best 95\% C.L. limit on tensor modes).}
\label{fig:result_comparison}
\end{figure*}

The bottom sub-panels in
Figure \ref{fig:result_summary} show the window and transfer functions
for each bin computed by pipeline A. Figure \ref{fig:maps} shows
the maps for patch CMB-1 computed by pipeline B, and
Figure \ref{fig:result_comparison} shows the QUIET power spectra in
comparison with the most relevant experiments in our multipole
range\footnote{Since \cite{larson:2010} does not provide an upper
limit on the BB power, we use the diagonal elements of the Fisher matrix
and show the points that are
$1.65\,\sigma$ above their central values as 95\% upper limits.}.
Additional plots
and data files are online\footnote{http://quiet.uchicago.edu/results/index.html}.

Fitting only a free amplitude, $q$, to the EE spectrum\footnote{Only
$\ell \ge 76$ are used in the EE fit and the $\chi^2$ calculation
relative to $\Lambda$CDM because the first EE bin has a significant
foreground contribution; see Section~\ref{sec:foregrounds}.} relative to the
7-year best-fit \textit{WMAP} $\Lambda$CDM spectrum \citep{larson:2010}, we
find $q = 0.87 \pm 0.10$ for pipeline A and $q = 0.94 \pm 0.09$ for
pipeline B. Taking into account the full non-Gaussian shapes of the
likelihood functions, both results correspond to more than a 10-$\sigma$
detection of EE power. In particular, in the region of the first peak, $76 \leq \ell \leq 175$, 
we detect EE polarization with more than 6-$\sigma$ significance, confirming the only other detection of this peak made by BICEP at higher frequencies.  
The $\chi^2$ relative to the  $\Lambda$CDM
model, with $C_\ell^{EB} = C_\ell^{BB} = 0$, is 31.6 (24.3) with 24 degrees of 
freedom, corresponding to a PTE of 14\% (45\%) for pipeline A (B).

\begin{table}[t]\renewcommand{\arraystretch}{1.5}
\begin{center}
\caption{CMB-Spectra Band Powers from QUIET Q-Band Data}
\vspace{-0.05in}
\begin{tabular}{ccccccc}
\hline
\hline
$\ell$ bin & EE & BB & EB \\
\hline

      25-75& \tablenotemark{a}$0.33^{ +0.16}_{ -0.11} $ & $  -0.01^{ +0.06}_{ -0.04} $
 & $  0.00^{ +0.07}_{ -0.07} $ \\
      76-125& $   0.82^{ +0.23}_{ -0.20} $ & $   0.04^{ +0.14}_{ -0.12} $
 & $  -0.10^{ +0.11}_{ -0.12} $ \\
     126-175& $   0.93^{ +0.34}_{ -0.31} $ & $   0.24^{ +0.28}_{ -0.25} $
 & $   0.71^{ +0.22}_{ -0.20} $ \\
     176-225& $   1.11^{ +0.58}_{ -0.52} $ & $   0.64^{ +0.53}_{ -0.46} $
 & $   0.18^{ +0.38}_{ -0.38} $ \\
     226-275& $   2.46^{ +1.10}_{ -0.99} $ & $   1.07^{ +0.98}_{ -0.86} $
 & $  -0.52^{ +0.68}_{ -0.69} $ \\
     276-325& $   8.2^{ +2.1}_{ -1.9} $ & $   0.8^{ +1.6}_{ -1.4} $
 & $   0.9^{ +1.3}_{ -1.3} $ \\
     326-375& $  11.5^{ +3.6}_{ -3.3} $ & $  -2.2^{ +2.7}_{ -2.4} $
 & $   0.0^{ +2.0}_{ -2.0} $ \\
     376-425& $  15.0^{ +6.2}_{ -5.8} $ & $  -4.9^{ +5.3}_{ -4.9} $
 & $   3.2^{ +3.9}_{ -3.9} $ \\
     426-475& $  21^{+13}_{-11} $ & $   2^{+11}_{-10} $
 & $   4.5^{ +8.3}_{ -8.2} $ \\

\hline
\end{tabular}
\label{tab:results}
\end{center}
\vspace{-0.15in}
\tablecomments{Units are thermodynamic temperatures, $\mu$K$^2$, scaled as
$C_\ell\ell(\ell+1)/2\pi$.} 
\tablenotetext{a}{Patch CMB-1 has significant foreground
contamination in the first EE bin.}
\end{table}

\begin{figure}[t]
\vspace{-0in}
\centering
\includegraphics[width=\linewidth]{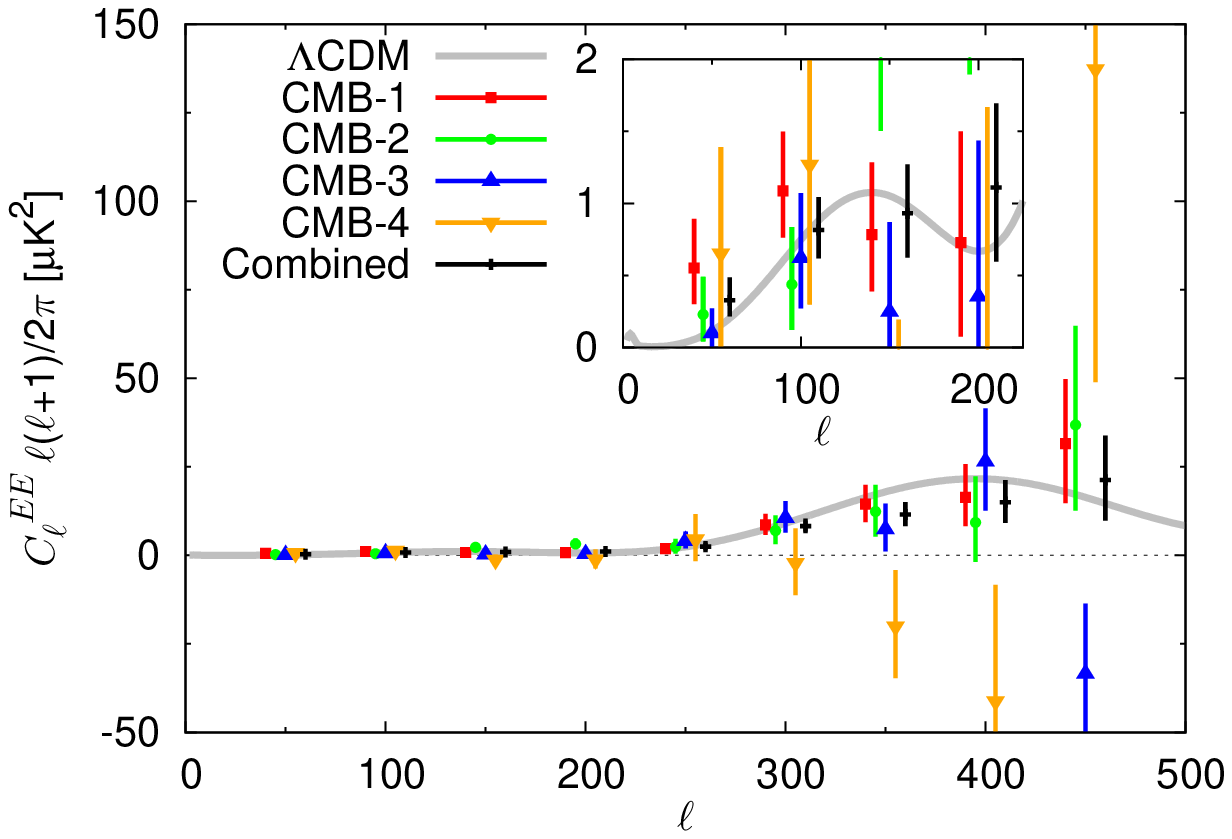}

 \vspace{-0.1in}

\includegraphics[width=\linewidth]{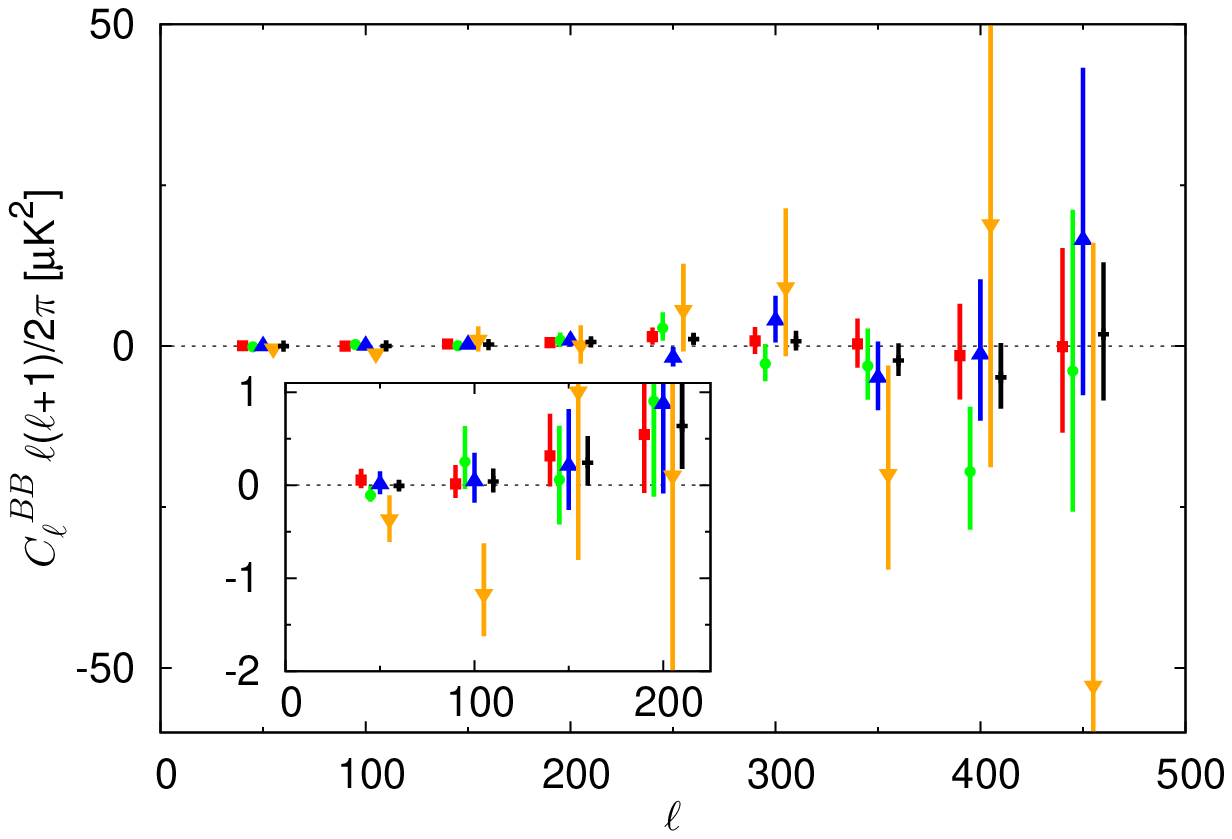}

 \vspace{-0.1in}

\caption{CMB power spectra are shown for each patch individually.  The top and bottom panels show the EE and BB spectra, respectively.  The different error bars for each patch mainly reflect the amounts of time each was observed.
}
\label{fig:patch_compare}
\end{figure}

\subsection{Foreground Analysis}
\label{sec:foregrounds}

In order to minimize possible foreground contamination, QUIET's four CMB patches were chosen to be far from the Galactic plane and known Galactic synchrotron spurs. In these regions, contributions from thermal dust emission are negligible in Q band. Spinning dust is expected to be polarized at no more than a few percent in Q band \citep{Battistelli2006, Lopez2010}, so we expect the contribution to polarized foreground emission in our patches to be small.  We therefore consider only two dominant sources of possible foreground contamination, namely compact radio sources and Galactic diffuse synchrotron emission.

To limit the effect of compact radio sources, we apply a compact-source
mask to our maps before computing the power spectra, as described in
Section~\ref{sec:analysis}.  We
also evaluate the CMB spectra both with and without the full
\textit{WMAP} temperature compact-source mask \citep{gold:2010}, and find
no statistically significant changes. 
The possible contribution from compact radio sources with fluxes below
the \textit{WMAP} detection level (1\,Jy) is small: 0.003\,$\mu$K$^2$  at $\ell=50$ and 0.01\,$\mu$K$^2$  at $\ell=100$ \citep{Battye2010}
\footnote{The estimate is robust since 90\% of the contribution
comes from the high-flux population between 100\,mJy and 1\,Jy, where
the distribution of the population is well understood.}.
We therefore conclude that our results are robust with respect to contamination from compact radio sources and that the dominant foreground contribution comes from diffuse synchrotron emission.

In Figure \ref{fig:patch_compare} we show the power spectra measured
from each patch. The CMB-1 EE band power for the first bin is $0.55\pm 0.14\,\mu\textrm{K}^2$,
a 3-$\sigma$ outlier relative to the expected $\Lambda$CDM band power of $0.13\,\mu\textrm{K}^2$; while not significant enough to spoil the overall agreement among the patches as shown in  Section~\ref{sec:nulls}, this is a candidate for a bin with foreground contamination.

To estimate the Q-band polarized synchrotron contamination in our CMB
patches, we process the \textit{WMAP}7 K-band (23-GHz) map through pipeline A
and estimate its band power, $\hat{C}^\mathrm{KK}_b$, as well as the
cross spectra with the QUIET Q-band data, $\hat{C}^\mathrm{QK}_b$.
These results are shown for the first bin ($25 \leq \ell \leq 75$; $b
=1$) in Table~\ref{tab:wmap_k_band_foreground_powers}, together with
the corresponding QUIET band powers, $\hat{C}^\mathrm{QQ}_b$. Since
foregrounds do not contribute to the sample variance, the
uncertainties for $\hat{C}^\mathrm{KK}_{b=1}$ and $\hat{C}^\mathrm{QK}_{b=1}$
are given by instrumental noise only, including contributions from
both \textit{WMAP} and QUIET. For $\hat{C}^\mathrm{QQ}_{b=1}$, 
sample variance as predicted by the $\Lambda$CDM model is also included.

There is significant EE power in patch CMB-1 
as measured by $\hat{C}^\mathrm{KK}_{b=1}$.  We also find a correspondingly significant 
cross correlation between the \textit{WMAP} K band and the QUIET 
Q band, confirming that this excess power is not due to 
systematic effects in either experiment and is very likely a foreground. No 
significant power is found in any other case.
The non-detection of foreground power at $\ell > 75$ is 
consistent with the expected foreground dependence: 
$\propto \ell^{-2.5}$ \citep{Carretti:2009gc}, and the low power found in $\hat{C}_{b=1}^\mathrm{KK}$.

The excess power observed in the first EE bin of CMB-1 is fully consistent
with a typical synchrotron frequency spectrum. To see this, we
extrapolate $\hat{C}^\mathrm{KK}_{b=1}$ from K band to Q band, assuming a
spectral index of $\beta=-3.1$ \citep{dunkley:2009}, and calculate the
expected power in $C^\mathrm{QK}_{b=1}$ and $C^\mathrm{QQ}_{b=1}$,
\begin{equation}
 C^\mathrm{QK}_{b=1}
  = \frac{1.05}{1.01} \left( \frac{43.1}{23} \right)^\beta \hat{C}^\mathrm{KK}_{b=1}
  = 2.57 \pm 0.69\,\mu\textrm{K}^2\:,
\end{equation}
\begin{equation}
 C^\mathrm{QQ}_{b=1}
  = \left[\frac{1.05}{1.01} \left( \frac{43.1}{23} \right)^\beta\right]^2 \hat{C}^\mathrm{KK}_{b=1}
  = 0.38 \pm 0.10\,\mu\textrm{K}^2\:,
\end{equation}
where the prefactor accounts for the fact that $\beta$ is defined in
units of antenna temperature, and the uncertainties are scaled from that
of $\hat{C}^\mathrm{KK}_{b=1}$.  These predictions are fully consistent with
the observed values of 
$\hat{C}^\mathrm{QK}_{b=1}$ and $\hat{C}^\mathrm{QQ}_{b=1}$, when combined with the $\Lambda$CDM-expected power.
We conclude that the
excess power is indeed due to synchrotron emission.

\begin{table}[htbp]
 \begin{center}
 \caption{ \label{tab:wmap_k_band_foreground_powers}
 Band and Cross Powers for $\ell=25$--75}

 \vspace{-0.15in}

 \begin{tabular}{clrrr}
  \hline
  \hline \\*[-8pt]
  Patch & Spectrum  & \multicolumn{1}{c}{$\hat{C}^\mathrm{KK}_{b=1}$} &
\multicolumn{1}{c}{$\hat{C}^\mathrm{QK}_{b=1}$} &
\multicolumn{1}{c}{$\hat{C}^\mathrm{QQ}_{b=1}$} \\*[+2pt]
  \hline
  CMB-1 & EE & $ \mathbf{ 17.4 \pm 4.7} $ &  $ \mathbf{ 3.30 \pm 0.55 } $  & $ \mathbf{ 0.55 \pm 0.14 } $ \\
        & BB &  $4.8 \pm 4.5$ &  $0.40 \pm 0.41$ & $ 0.06 \pm 0.08$ \\
        & EB & $-6.2 \pm 3.2$ &  $0.27 \pm 0.38$ & $ 0.10 \pm 0.08$ \\
\hline
  CMB-2 & EE &  $5.5 \pm 3.7$ &  $0.01 \pm 0.56$ & $ 0.23 \pm 0.19$ \\
        & BB &  $4.6 \pm 3.4$ &  $0.18 \pm 0.48$ & $-0.11 \pm 0.13$ \\
        & EB & $-5.5 \pm 2.8$ & $-0.39 \pm 0.41$ & $-0.20 \pm 0.12$ \\
\hline
  CMB-3 & EE &  $0.2 \pm 1.9$ &  $0.64 \pm 0.43$ & $ 0.10 \pm 0.18$ \\
        & BB & $-0.3 \pm 2.6$ &  $0.33 \pm 0.35$ & $ 0.01 \pm 0.13$ \\
        & EB &  $1.4 \pm 1.7$ & $-0.34 \pm 0.30$ & $ \mathbf{ -0.27 \pm 0.11}$ \\
\hline
  CMB-4 & EE & $-5.2 \pm 5.1$ &    $0.7 \pm 1.2$ & $ 0.65 \pm 0.58$ \\
        & BB & $-2.6 \pm 5.2$ &   $-0.1 \pm 1.1$ & $-0.37 \pm 0.52$ \\
        & EB & $-1.0 \pm 3.9$ &    $0.0 \pm 0.9$ & $-0.15 \pm 0.47$ \\
  \hline 
 \end{tabular}
 \\

 \end{center}

 \vspace{-0.1in}

\tablecomments{
 Power-spectra estimates for the first multipole bin for each
 patch, computed from the \textit{WMAP}7 K-band data and the QUIET Q-band
 data. The units are $\ell (\ell+1) C_\ell / 2 \pi~(\mu \mathrm{K}^2)$
 in thermodynamic temperature. Uncertainties for
 $\hat{C}^\mathrm{KK}_{b=1}$ and $\hat{C}^\mathrm{QK}_{b=1}$ include noise
 only. For $\hat{C}^\mathrm{QQ}_{b=1}$ they additionally include CMB
 sample variance as predicted by $\Lambda$CDM. Values in bold are more than $2\,\sigma$ away from zero.}
  
\end{table}

\subsection{Constraints on Primordial B modes}
\label{subsec:r}

We constrain the
tensor-to-scalar ratio, $r$, using the QUIET measurement of the BB power spectrum at low multipoles ($25 \leq \ell \leq 175$).  Here $r$ is defined as the
ratio of the primordial--gravitational-wave amplitude to the curvature-perturbation amplitude at a scale $k_0=0.002$\,Mpc$^{-1}$.  We then fit
our measurement to a BB-spectrum template computed from the $\Lambda$CDM
concordance parameters with $r$ allowed to vary.  For simplicity, we
fix the tensor spectral index at $n_t=0$ in computing the
template\footnote{Our definition of $r$ agrees
with \cite{Chiang:2010}}.  This choice makes the BB--power-spectrum
amplitude directly proportional to $r$.

For pipeline A, we find $r=0.35^{+1.06}_{-0.87}$, corresponding to $r
< 2.2$ at 95\% confidence. Pipeline B obtains 
$r=0.52^{+0.97}_{-0.81}$.   The results are consistent;  the lower panel of Figure  
\ref{fig:result_comparison} shows our limits on BB power  in comparison with those from BICEP, QUaD, and \textit{WMAP}.  QUIET lies between BICEP and \textit{WMAP} in significantly limiting $r$ from measurements of CMB--B-mode power in our multipole range.
Although we neither expected nor detected any BB
foreground power, the detection of an EE foreground in patch CMB-1 suggests that
BB foregrounds might be present at a smaller level.  We emphasize that 
the upper limit we report  is therefore conservative.

\subsection{Temperature Power Spectra}
\label{subsec:temperature}

\begin{figure}[!]
\vspace{-0in}
\centering
\includegraphics[width=\linewidth]{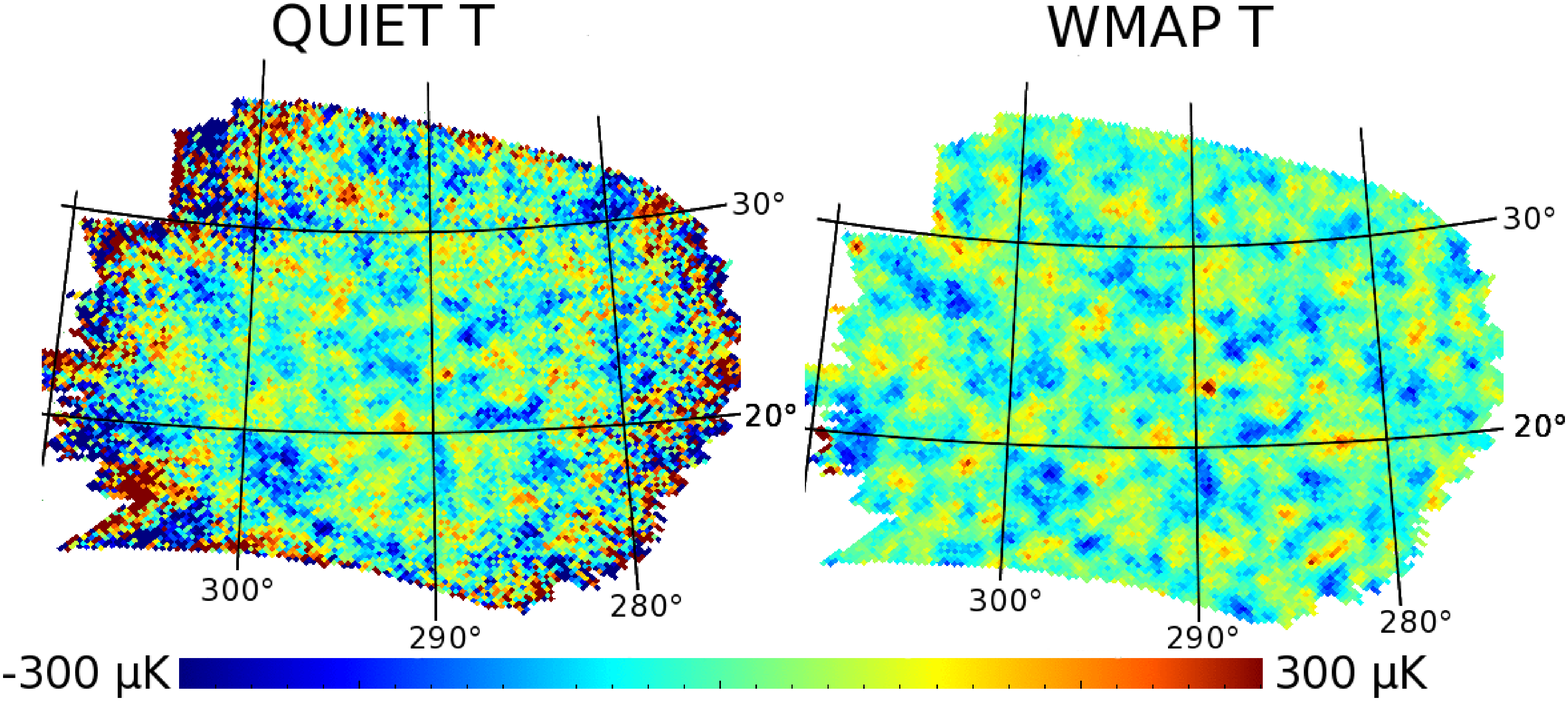}
\includegraphics[width=\linewidth]{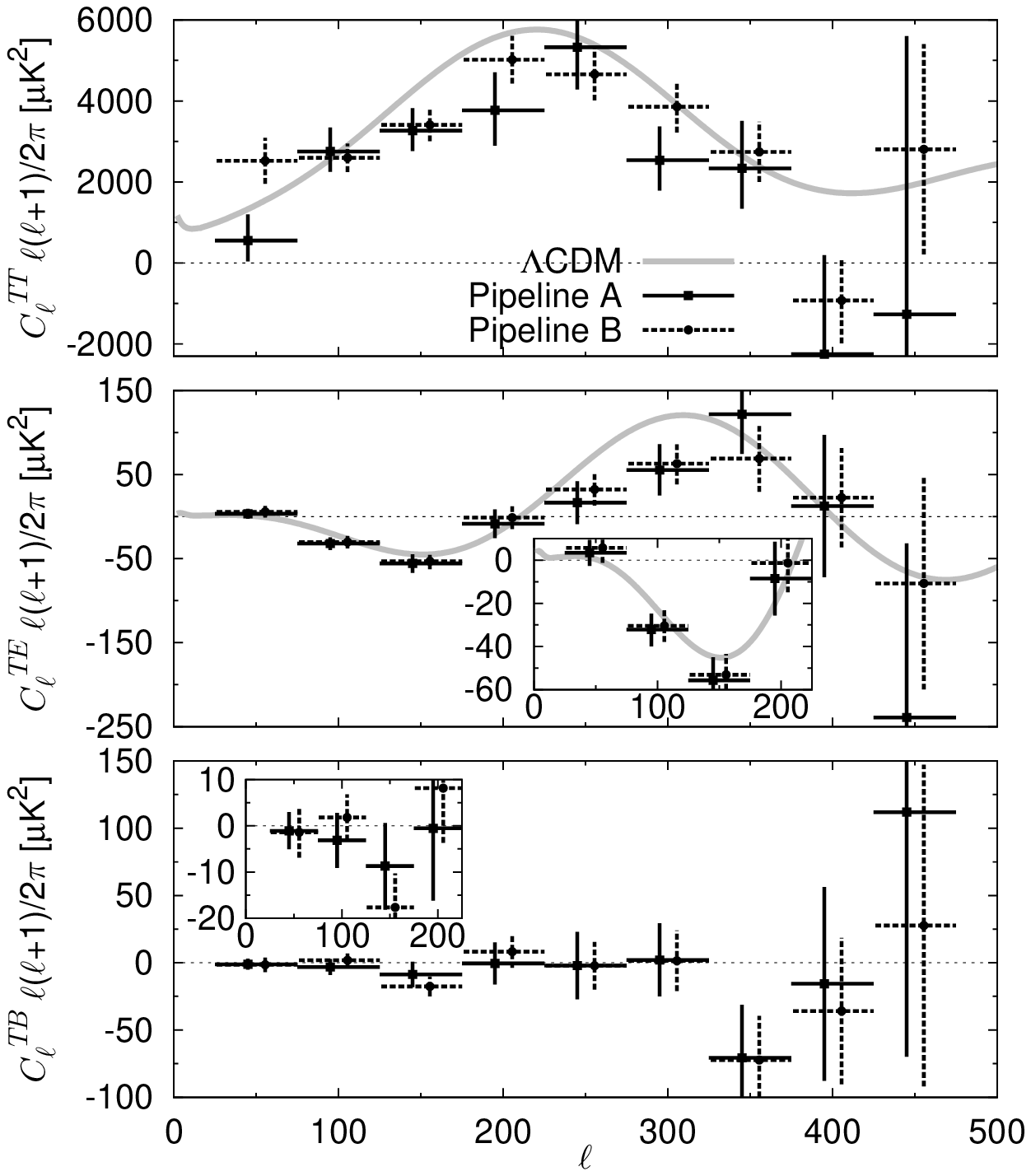}

 \vspace{-0.2in}

\caption{The top row compares our temperature map to the \textit{WMAP} 7-year Q-band map \citep{Jarosik:2010iu} for patch CMB-1 in Galactic coordinates.   Lower panels show the CMB temperature power spectra: TT, TE, and TB.}
\label{fig:results_T}
\end{figure}

 Figure \ref{fig:results_T} compares the QUIET and \textit{WMAP} Q-band temperature maps and TT, TE, and TB power spectra.
Agreement with the $\Lambda$CDM model is good.
This is a strong demonstration of the raw sensitivity of the QUIET
detectors; the single QUIET differential-temperature assembly produces a high--signal-to-noise map using only 189\,hours (after selection) of observations. 
The high sensitivity of these modules makes them very useful for
calibration, pointing estimation, and consistency checks (see
Section \ref{sec:calibration}).

 \vspace{0.3in}

\section{Systematic Errors}
\label{sec:systematics}

\begin{figure*}[htbp]
\centering
\includegraphics[width=6.8in]{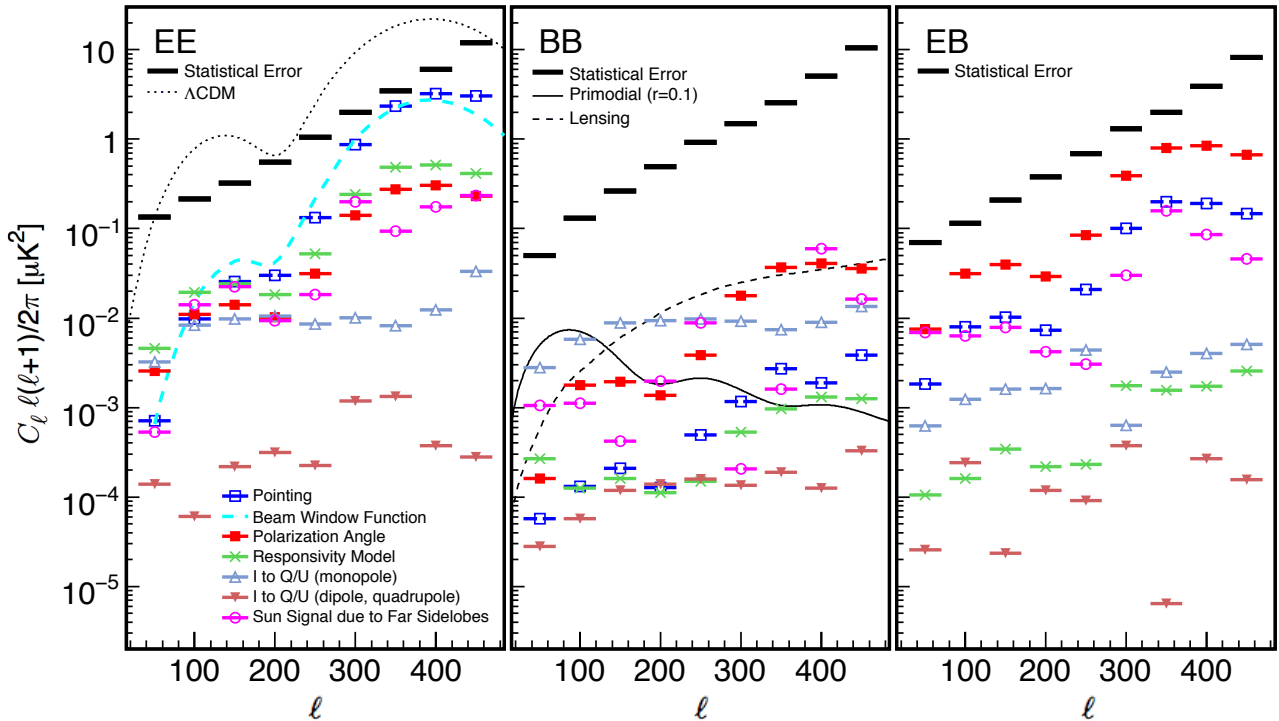}
\caption{Systematic uncertainty estimates for EE, BB, and EB power
 spectra.  Estimates for a variety of effects (see text) are shown for
 the three power spectra.  In all cases, they are well below the statistical errors, which are also shown.  In particular, the contaminations to the primordial--B-mode signal, at multipoles below 100, are below the level of $r = 0.1$, even though we do not make a correction for the largest contaminant,  the monopole leakage. }
\label{fig:syst_summary}
\end{figure*}

The passing of the null suite itself limits systematic uncertainty,  but to get well below the statistical errors, dedicated studies are needed. They are important in gaining confidence in the result and also in evaluating the potential of the methods and techniques we use for future efforts.
We pay special attention to effects that can generate false B-mode
signals.
Our methodology is to simulate and then propagate calibration
uncertainties (see Section~\ref{sec:calibration}) and other 
systematic effects through the entire pipeline.
The systematic errors in the power spectra are shown in Figure~\ref{fig:syst_summary}.  The possible contaminations are well below the statistical errors;  in particular, the levels of spurious B modes are less than the signal of $r=0.1$.  This is the lowest level of BB contamination yet reported by any CMB experiment.  This section describes how each effect in  Figure~\ref{fig:syst_summary} is determined and 
considers three additional possible sources of contamination.

An uncertainty not shown in Figure~\ref{fig:syst_summary} is that arising from
the 
overall responsivity error estimate of 6\% (12\% in power-spectra units).  After including the effect of possible time-dependent responsivity variations (4\%, see below), the power-spectra uncertainty is 13\%.  It is multiplicative, affecting all power-spectra results independent of multipole.

\subsection{Beam Window Function and Pointing} \label{sec:beam_sys}

The uncertainty in the beam window function is another multiplicative
factor, one which increases with multipole. 
We estimate this uncertainty using the difference
of the beam window functions measured for the central module and the modules
of the differential-temperature assembly, which are at the edge of the
array.
The difference is statistically significant, coming from the different locations (with respect to the optics) in the focal plane; it is expected from the pre-season antenna range measurements. 

Uncertainties in pointing lead to distortions in polarization maps. E
power will be underestimated and spurious B power (if the distortions are non-linear)  generated \citep{Hu/etal}. 
We quantify these effects by using the differences in pointing
solutions from two independent models: the fiducial model used for the
analysis and an alternative model based on a different set of 
calibrating observations.
We also modeled and included the effects of the deck-angle--encoder shift which occurred for a portion of the season (Section~\ref{sec:cal_pol_angle}).

 \vspace{0.3in}

\subsection{Responsivity  and Polarization Angle}  \label{sec:gain_sys}

Responsivity shifts, particularly within CESes, lead to distortions
in the maps.  Full-pipeline simulations quantify the shifts
caused by 
variations in the cryostat or electronics temperatures.
Similarly shifts from using responsivities determined from the Moon
data, Tau~A data, or from the sparse wire grid, rather than those from
the sky dips, are determined.  
We also incorporate the uncertainty in the atmospheric-temperature
model used in analyzing the sky-dip data.
The largest possible effects on the power spectra are shown in
 Figure~\ref{fig:syst_summary}.

Uncertainties in the orientation of the polarization axes of the modules
can lead to leakage between E and B modes.  To quantify this leakage, we
use the differences in power spectra where these angles are determined
from Moon data, Tau~A data, and the sparse--wire-grid data.  As expected, the largest effects show up in EB power.
We also estimate systematic error due to possible fluctuation of the
detector angles over the course of the season.  The contribution is
negligibly small compared to the overall shift of the angles described
above.  Both effects are included in the ``Polarization Angle'' points
in Figure~\ref{fig:syst_summary}.

\subsection{Instrumental Polarization}
\label{sec:systematics_ip}
As described in Section~\ref{sec:calibration_ip}, the I to Q~(U) leakage coefficients for the QUIET detector diodes are small:  1\%~(0.2\%).  
Except in the case of patch CMB-4, our scanning strategy significantly reduces this effect with the combination of sky and deck-angle rotation.

We estimate spurious $Q$ and $U$ in the maps for each CES-diode using the
\textit{WMAP} temperature map and our known leakages.
Shown in Figure~\ref{fig:syst_summary} are the estimates of spurious EE,
BB, and EB powers from full-pipeline simulations, where
for each realization
the spurious $Q$ and $U$ are added to the $Q$ and $U$ from simulated $\Lambda$CDM
E modes.
While this method has an advantage of being able to use the real
(not simulated) temperature map, it does not incorporate TE correlation,
which only affects the spurious EE power.
As a complement, we repeat the study, but using simulated $\Lambda$CDM
maps for both temperature and polarization; this only changes the estimate of
spurious EE power by 30\% at most.
Because the spurious power is as small as it is, 
we have treated it as a systematic rather than correcting for it.  
Doing so would give us a further order of magnitude suppression.

Differing beam ellipticities can also induce higher multipole
polarization signals.  We measure these leakages from Tau~A and Jupiter
observations and find that the higher-order multipoles are at most 0.1\%
of the main-beam peak amplitude.
The corresponding effects on the power spectra, which are seen in Figure~\ref{fig:syst_summary}, are of little concern.

\subsection{Far Sidelobes Seeing the Sun}
\label{sec:systematics_sun}
While we make cuts to reduce the effects of far sidelobes seeing the Sun
(Sections \ref{sec:instrument} and \ref{sec:data:tod:filter}),  small contaminations could remain.  We make full-season
maps for each diode in Sun-centered coordinates and then use these
maps to add contamination to full-pipeline CMB simulations.
The excess power found in the simulations is taken as the systematic uncertainty.
We do not observe any signature from the Moon, nor do we cut on
proximity to it.  We estimate the related systematic error and find
that it is negligibly small compared to that assigned to the
contamination from the Sun.

\subsection{Other Possible Sources of Systematic Uncertainty}
\label{sec:systematics_other}
Here we discuss a few additional potential sources of systematic uncertainty, which are found to be subdominant.

{\bf Ground-Synchronous Signals.} 
QUIET's far sidelobes do see the ground for some diodes at particular elevations and deck angles.  Ground pickup that is constant throughout a CES is removed by
our TOD filters;  the net effect of this filtering in the full-season maps
is a correction of $\approx 1\,\mu\mathrm{K}$.

The only concern is ground pickup that changes over the short span
of a single CES.  
We find little evidence for changes even over the entire season, let alone over a single CES.  We therefore conservatively place an upper limit on such changes  using the statistical errors on the ground-synchronous signal.
We start with the CES and module with the largest ground pickup.  We then simulate one day's worth of data, inserting a ground-synchronous signal that changes by its statistical error.
Given the distribution in the magnitude of the ground-synchronous signal and assuming that changes in this signal are proportional to the size of the signal itself, by considering that the signals from changing pickup add incoherently  into the maps made from multiple CES-diodes at a variety of elevations and deck angles, we estimate
an upper limit on residual B power from possible changing ground-pickup signals.  The result is $\lesssim 10^{-4}\,\mu \mathrm{K}^2$ at multipoles below 100.

{\bf ADC Non-linearities.} The possible residual after the correction for the non-linearity in the ADC system results in effects similar to the I to Q (or U) leakage and the variation of the responsivity during the CES.
We estimate such effects based on the uncertainty in the correction parameters,
confirming that there is at most a 3\% additional effect for the leakage
bias, 
and that the responsivity effect is also small, less than half of
the systematic error shown for the responsivity in Figure~\ref{fig:syst_summary}.

{\bf Data-Selection Biases.} Cuts can cause biases if they are, for example, too stringent.  We expect none but to be sure we apply our selection criteria to 144 CMB + noise simulations.  No bias is seen, and in particular we limit any possible spurious B modes from this source 
to $\lesssim10^{-3}\,\mu\mathrm{K}^2$ at multipoles below 100.

\section{Conclusions}
\label{sec:conclusions}

QUIET detects polarization in the EE power spectrum at 43\,GHz.
We confirm with high significance the  detection of polarization in the region of the
first acoustic peak \citep{Chiang:2010} in the multipole region $\ell = 76$--175.  We find no significant power in
 either BB or EB  between
$\ell=25$ and $\ell=475$. We measure the tensor-to-scalar
ratio to be $r=0.35^{+1.06}_{-0.87}$.

These results are supported by a very extensive suite of null tests in
which  42 divisions of data were used for each of 33
different cut configurations. The
selection criteria and systematic errors were determined before
the power spectra themselves were examined. Biases were revealed
during this process, the last of which was a contamination
present in the null spectra at the level of about $20\%$ of the
statistical errors, but eliminated when cross-correlating maps
with differing telescope pointings. The robustness of the final
results is further supported by having two pipelines with results in
excellent agreement, even though one
uses only cross correlations while the other also uses
auto correlations.

Several possible systematic effects are studied with full end-to-end
simulations.  The possible contaminations in the B-mode power are thereby limited to
a level smaller than for any other published experiment:  below the level
of $r = 0.1$ for the primordial B modes; simply correcting for the known level of
instrumental
polarization would reduce this to $r < 0.03$.  This very low level of
systematic uncertainty comes from the combination of
several important design features, including a new time-stream
``double-demodulation'' technique, side-fed Dragonian optics, natural sky
rotation, and frequent deck rotation.

The correlation modules we use have a polarization sensitivity (Q and U
combined) of $280\,\mu\textrm{K}\sqrt{\textrm{s}}$, leading to an array
sensitivity of 69\,$\mu\textrm{K}\sqrt{\textrm{s}}$.  Further, the
$1/f$ noise observed in our detectors is small:
the median  knee frequency is just 5.5\,mHz. One important
outcome of this work, then, is the demonstration that our detectors,
observing from a mid-latitude site, give excellent sensitivity and systematic
immunity.

Because of our mid-latitude site, we are driven to collect data in four separate patches.  While we lose some sensitivity (compared to going deeper on a single patch), there are a few advantages that we have exploited.  The patches are scanned differently, in terms of time of day and the degree of crosslinking, and these differences allow some important systematic checks.  Another advantage concerns foregrounds.  

Foreground contamination is expected to be one of the main
limiting factors in the search for primordial B modes.
Indeed we report a  3-$\sigma$ detection of  synchrotron emission in one
of our four CMB patches,  originally chosen for their expected
low foreground levels.
Our detection is only in EE; our BB 2-$\sigma$ limit permits a BB signal
about half as large. If we extrapolate that value to the foreground minimum of
about 95\,GHz,
we would have synchrotron contamination at the level of $r = 0.05$ from this one
patch. Neither \textit{WMAP}
nor \textit{Planck} will have enough sensitivity~\citep{Tauber2010} to sufficiently
constrain the polarized synchrotron
amplitude at this level in any patch. In fact, our Q-band polarization maps
are already as deep or deeper than
what \textit{Planck} will achieve at the same frequency.
Dedicated
low-frequency observations are clearly needed to achieve such constraints.  When foreground cleaning becomes important, consistency among  separate patches will be an important handle on our understanding.

Further progress must be made through larger arrays and
longer integration times. In hand we have data collected by
the 90-element W-band array with similar sensitivity to our Q-band
array and more than twice the number of observing hours. Results from the analysis of that data set
will be reported in future publications. A W-band receiver with the sensitivity to reach below the level of $r=0.01$ is under development.

\acknowledgements
Support for the QUIET instrument and operation comes through the NSF
cooperative agreement AST-0506648. Support was also provided by NSF awards
PHY-0355328, AST-0448909, AST-1010016, and PHY-0551142; KAKENHI 20244041,
20740158, and 21111002; PRODEX C90284; a KIPAC Enterprise grant; and by the Strategic Alliance for
the Implementation of New Technologies (SAINT).

Some work was performed on the Joint Fermilab-KICP Supercomputing
Cluster, supported by grants from Fermilab, the Kavli Institute for
Cosmological Physics, and the University of Chicago.  Some work was
performed on the Titan Cluster, owned and maintained by the University
of Oslo and NOTUR (the Norwegian High Performance Computing
Consortium), and on the Central Computing System, owned and operated
by the Computing Research Center at KEK.  Portions of this work were
performed at the Jet Propulsion Laboratory (JPL) and California
Institute of Technology, operating under a contract with the National
Aeronautics and Space Administration. The Q-band polarimeter modules
were developed using funding from the JPL R\&TD program.

C.D. acknowledges an STFC Advanced Fellowship and an ERC IRG grant under FP7.
P.G.F. and J.A.Z. gratefully acknowledge the support of the Beecroft Institute of Particle Astrophysics and Cosmology, the Oxford Martin School, and the Science and Technology Facilities Council.
L.B., R.B., and J.M. acknowledge support from CONICYT project Basal PFB-06.
R.B. acknowledges support from ALMA-Conicyt 31080022 and 31070015.
A.D.M. acknowledges a Sloan foundation fellowship.

PWV measurements were provided by the Atacama Pathfinder Experiment
(APEX). We thank CONICYT for granting permission to operate within the
Chajnantor Scientific Preserve in Chile, and ALMA for providing site
infrastructure support.
Field operations were based at the Don Esteban facility run by Astro-Norte.
We are particularly indebted to the engineers
and technician who maintained and operated the telescope: Jos\'e Cort\'es,
Cristobal Jara, Freddy Mu\~noz, and Carlos Verdugo.

In addition, we would like to acknowledge the following people for
their assistance in the instrument design, construction,
commissioning, operation, and in data analysis: Augusto Gutierrez
Aitken, Colin Baines, Phil Bannister, Hannah Barker, Matthew R. Becker, Alex
Blein, Mircea Bogdan, April Campbell, Anushya Chandra, Sea Moon Cho,
Emma Curry, Maire Daly, Richard Davis, Fritz Dejongh, Joy Didier, Greg
Dooley, Hans Eide, Will Grainger, Jonathon Goh, Peter Hamlington,
Takeo Higuchi, Seth Hillbrand, Christian Holler,
Ben Hooberman, Kathryn D. Huff, William
Imbriale, Koji Ishidoshiro, Eiichiro Komatsu,
Jostein Kristiansen, Richard Lai, Erik Leitch, Kelly
Lepo, Martha Malin, Mark McCulloch, Oliver Montes, David Moore, Makoto
Nagai, Ian O'Dwyer, Stephen Osborne, Stephen Padin,
Felipe Pedreros,
Ashley Perko, Alan Robinson,
Jacklyn Sanders, Dale Sanford, Mary Soria, Alex Sugarbaker, 
David Sutton,
Matias Vidal, Liza Volkova, Edward Wollack, Stephanie Xenos, and Mark
Zaskowski.

\end{document}